\documentclass[aps,prd,preprint,showpacs,showkeys,nofootinbib,
longbibliography,12pt]{revtex4-1}

\usepackage{amssymb,amsmath}

\usepackage{graphicx,color}
\usepackage[utf8]{inputenc}
\def\be{\begin{eqnarray} &&}
\def\nonu{\nonumber \\ &&}
\def\nonub{\right. \nonumber \\ && \left.}
\def\ee{\end{eqnarray}}
\def\psla{\slash \! \! \!}

\newcommand{\bmm}[1] {\mbox{\boldmath{$#1$}}}
\newcommand {\spro}[2]{(#1 \cdot #2)}
\begin{document}

\title{Solving the  Bethe-Salpeter Equation in Minkowski Space for a 
Fermion-Scalar system}


\author{J.H.  Alvarenga Nogueira} 
\affiliation{Instituto Tecnol\'ogico de Aeron\'autica \& Universit\`a di Roma 'La Sapienza'
\& Istituto Nazionale di Fisica Nucleare, Sezione di Roma, Piazzale A. Moro 2, I-00185 Roma, Italy}
\email{dealvare@roma1.infn.it}
\author{V. Gherardi}
\affiliation{SISSA International School for Advanced Studies \& Istituto Nazionale di Fisica Nucleare, Sezione di Trieste,
 Via Bonomea 265, 34136, Italy\footnote{ Research
 thesis in fulfillment of the requirements for the master degrees at the  University ''La Sapienza'', Rome, Italy}}
 \email{vgherard@sissa.it}
\author{T. Frederico} 
\affiliation{Instituto Tecnol\'ogico de Aeron\'autica, DCTA, 12228-900 S\~ao Jos\'e dos Campos,~Brazil}
\email{tobias@ita.br}

\author{G. Salm\`e}
\affiliation{Istituto Nazionale di Fisica Nucleare, Sezione di Roma, Piazzale A. Moro 2, I-00185 Roma, Italy}
\email{salmeg@roma1.infn.it}

\author{D. Colasante, E. Pace}
\affiliation{ Universit\`a di Roma 'Tor Vergata' and
    Istituto Nazionale di Fisica Nucleare, Sezione di Roma
    Tor Vergata,
  Via della Ricerca Scientifica 1, I-00133, Rome,  Italy}
\email{emanuele.pace@roma2.infn.it}

\begin{abstract}
The ladder Bethe-Salpeter Equation of a bound $(1/2)^+$ system, 
composed by a fermion and 
a scalar boson,
is solved  in Minkowski space, for the first time. 
The formal tools are the same already
successfully adopted for two-scalar and two-fermion systems, namely the Nakanishi
integral representation of the Bethe-Salpeter amplitude and the  
light-front
projection of the fulfilled equation. Numerical results are presented and  
discussed for two
interaction kernels: i) a massive scalar exchange and
ii) a massive vector exchange, illustrating both the correlation
between  
binding energies and the interaction coupling constants, as well as the valence content of
the interacting state, through the valence probabilities and the light-front
momentum distributions. In the case of the scalar
exchange, an interesting side effect, to be ascribed to the repulsion generated 
by the 
small components of the Dirac spinor,  is pointed out, { while for the vector
exchange the manifestation of the helicity conservation opens new interesting 
questions to be addressed 
within a fully non-perturbative  framework, as well as the onset of a
scale-invariant regime.}
\end{abstract}

\maketitle


\section{Introduction}
Within the relativistic quantum field theory, the intrinsically 
non-perturbative nature of a bound system can be  suitably treated by an
integral equation, like
the homogeneous Bethe-Salpeter equation (BSE) \cite{BS51}. {  As it is well-known,  
the path-integral approach on the Euclidean lattice is the main tool for addressing the non
perturbative regime.   However,  
efforts  to get actual solutions of the
BSE in Minkowski
space, where the physical processes take place, are highly
desirable.
 In order to carry out in its full glory the program of constructing a {\em continuous } approach, able to yield a phenomenological
 description of    the dynamics inside   bound systems in
Minkowski space, one has to consider fundamental ingredients in the BSE, that  
make us immediately understand the big 
challenge  to be faced.
Schematically,  main ingredients in the BSE are in order; i) the dressed
propagators of the constituents and quanta, ii) the fully dressed interaction kernel,  constructed from
the two-particle-irreducible diagrams. This means that in
order to get a {refined} description of the dynamics inside a composite
relativistic system one cannot limit to consider the BSE, but one has to widen
the framework including the Dyson-Schwinger equations (DSEs) for the self-energies
and, in principle, an infinite tower of DSEs that consistently 
determine the above
ingredients.  In view of establishing  a workable set of integral
equations for studying the adopted   Lagrangian, a coherent 
truncation scheme of the aforementioned infinite DSE tower, 
 but able to preserve
the  symmetries dictated by the
investigated interaction, is a prerequisite, {
(see, e.g., Ref. \cite{Binosi:2016rxz}, for a recent study of the issue, and
references quoted therein). Since more than two decades, the Euclidean space has been the elective one
where successful efforts have been carried out for developing the above sketched framework,
 based on both BSE 
and DSEs for
 the self-energies, with a
 well-controlled set of approximations, like the so-called rainbow-ladder 
 approximation. {Starting from the seminal review \cite{Roberts:1994dr}, 
 where the general framework  was illustrated by presenting the formalism
 for both QED
 and QCD together with  some first results, very soon the applications to hadron physics 
 became more and more
 sophisticated. As a matter of fact, the main features of 
 the non-Abelian gauge theory of the strong interaction, like confinement and 
 dynamical symmetry breaking, were addressed in an extended way (see, e.g.,
  Refs.
 \cite{Alkofer:2000wg,Maris:2003vk}),
 providing 
   constant improvements
 in the description of hadron observables, like mesonic and baryonic spectra 
 as well as  electromagnetic
 properties in the space-like region, directly addressable  in the Euclidean
 space (see, e.g., Refs.
 \cite{Tandy:2003hn,Fischer:2006ub,Bashir:2012fs,Eichmann:2016yit}).
 The interested reader can straightforwardly realize the huge amount of
  improvements
 reached
  in both  formalism and 
  obtained results  by the continuous approach in the Euclidean space
    (a recent introduction to the numerical methods can be found
   in Ref. \cite{Sanchis-Alepuz:2017jjd}) and appreciate the attempts to extend
    the BSE+DSE framework to the Minkowski space, for investigating 
     light-like and time-like quantities (considering also a due cross-check 
    for the
   space-like  ones)  } }
 
 { On the Minkowski side,
though   the necessity of elaborating a similar
framework is universally recognized (see, e.g., Refs. \cite{Roberts:1994dr,Eichmann:2016yit}, just to
mention
reviews well-distant in time) } one has a rather 
elementary  stage in the development, 
basically  { i) one does not take into accout 
self-energies and vertex
corrections and ii) considers interaction kernels mainly in ladder  
 approximation (at most with cross-ladder contributions)}.
 Nonetheless non trivial results can  be achieved,
 as briefly illustrated below,
particularly with regard to  the evaluation of both longitudinal and transverse light-front momentum 
distributions, { not directly addressable within a Euclidean framework without introducing ad 
hoc
paths {to be carefully treated}, like, e.g., analytic continuations (with all the well-known 
caveats about the singularities in the complex plane) or re-summation of 
infinite Mellin moments.} In order to
bring the Minkowskian approach to the level of sophistication of  the Euclidean one, we need  to build a systematic study of systems with
different degrees of freedom, so that we can gain the physical intuition
useful for guiding the next step, i.e.
the application of the approach to the gap equation (as some
groups are  elaborating, e.g., Refs. \cite{Sauli:2002tk,Mezrag}). Indeed, one could even devise   an intermediate step,
helpful for phenomenological applications, by adopting   the proposal contained in
Ref. \cite{Mello:2017mor}, where pion observables were evaluated
by using a Bethe-Salpeter (BS)  amplitude in Minkowski space (notice that in the quoted work an Ansatz
was considered),  together with a dressed quark propagator extracted from    lattice
data.
 
In this work, we  illustrate 
how to solve the homogeneous BSE, in ladder approximation  without vertex and self-energy
corrections, for a fermion-scalar 
system with positive parity, i.e. with quantum numbers $J^\pi=(1 / 2)^+$,
directly in Minkowski space.  Indeed, the achievements illustrated in what
follows are part of a more general 
investigation of the BSE in
Minkowski space, carried out within  an approach based on (i)
the so-called Nakanishi integral representation (NIR) of the 
BS amplitude (see, e.g., Ref. \cite{nak71} for the  general presentation of
the framework
 applied to  the n-leg transition amplitudes) and (ii) the
light-front (LF)
projection of the BSE,  i.e. its  restriction to a
vanishing relative LF time 
(an introduction to this technique and its application to the BSE is given
in \cite{FSV1}). This  approach, together with the analogous one developed by Carbonell and Karmanov 
(see, e.g.,
\cite{CK2006,CK2006b,CK2010}), has already achieved relevant outcomes, addressing: 
i) two-scalar systems  both in bound states
and at the zero-energy limit \cite{FSV1,FSV2,FSV3,Tomio2016}; the two-fermion bound state
 in a $0^+$ channel \cite{dFSV1,dFSV2}. 
 To summarize the results of the previous studies and of the present one, we can state that
  such an approach
is able to yield actual solutions of the BSE,
 directly in Minkowski space. Therefore one can be  confident to 
 reach a consistent (with  the set of  assumptions 
 discussed in what follows) 
 and reliable description of the inner dynamics of
 relativistic systems,  { such that it becomes feasible the evaluation of 
 relevant quantities, like  
 the 
 valence
 component of the system}.
  Within the BSE approach, 
  one can obtain a non-perturbative description
 of the  dynamics inside the system, in a space endowed with a $SO(3,1)$ 
 symmetry,  since an integral equation is able to sum up all the {\em infinite}
 contributions generated by the interaction kernel,  { though truncated at a given
 order in the coupling constant } Noteworthy, there are 
 efforts 
 to go beyond the ladder approximation by including the 
 cross-ladder diagrams as shown in  Refs. \cite{CK2006b,karmanov2006bethe,gigante2017bound}, 
 as well as to explore the formal inversion of
 the NIR, as in Ref. \cite{Carbonell:2017kqa}, for eventually exploiting a 
 Wick-rotated formulation of the BSE. 
 { It should be pointed out that  the inversion is an ill-posed problem that
 needs non trivial elaborations  to be accomplished. For instance,
   in Ref.  \cite{Chang:2013pq} (for a recent  Bayesian approach to the inversion, see
 Ref. \cite{Gao:2016jka}), the challenge of the inversion has been well illustrated, 
  showing how the pion parton distribution function
   could be evaluated     starting from
    a Euclidean framework where both the  quark-antiquark BSE and the quark gap-equation 
    are taken into account.}

{  The target of our investigation is the $(1/2)^+$ bound system, composed by
 a fermion and a scalar. 
 {As a prototype of such a system one could consider a {\em mock } nucleon 
 composed by a quark and a {\em point-like} scalar diquark (see, e.g., Ref. \cite{Alkofer:2000wg,Eichmann:2016yit}
  and references quoted therein for a general introduction to the  description 
  of   {a 
  baryon in terms of  confined  quark and  {\em extended}
 diquarks, with the last feature needed for implementing the correct statistics})}, or even a more exotic bound
system as the ghost-quark one investigated, e.g. in Ref. \cite{Alkofer:2011pe}.}
In order to broad our study, we allowed 
the constituents  to interact through two possible interaction Lagrangians: i) ${\cal L}=
 \lambda^s_F \bar \psi \psi \chi + \lambda^s_S \phi^*\phi \chi$ and 
 ii) ${\cal L}=
 \lambda^v_F \bar \psi \psla V\psi  - i\lambda^v_S \phi^*\overleftrightarrow{\partial}_\mu\phi V^\mu$, where
   only the coupling constant $\lambda^s_S$ has a mass dimension,
    while the other three couplings are dimensionless. The fields $\psi$ and
     $\phi$ describe the fermionic and bosonic constituents, respectively, 
     while
     $\chi$ and $V^\mu$ are the fields  of the exchanged scalar and  vector
      bosons. {  It is worth noticing that  in the mock nucleon 
      (representing only a first step in the avenue for
       developing a Minkowskian approach for investigating an actual  nucleon), 
      the only explicit vector boson exchange is
      between the quark and the point-like diquark, while in the
       modern approach the nucleon is bound
      by a quark exchange and the gluon exchange is buried in the interaction kernel
       \cite{Buck:1992wz,Alkofer:2000wg,Eichmann:2016yit}.
       Hence},
       at the present stage,   
      one can obtain the description of a massive quark-diquark system
only in the region dominated by the one-gluon exchange as it is discussed 
in Sect. \ref{sect_numres}, while for
the massless ghost-quark bound system it is necessary at least to dress the interaction in order to
break the scale invariance that the bare vertex brings about.
     
The BS amplitude for the system we are addressing is
\be
 \Phi^\pi(k,p,J_z)=\int {d^4 x} ~e^{ik\cdot x}~\langle 0| T\{\psi(x/2)~\phi(-x/2)\}|p;J,J_z,\pi\rangle
 ~~,\ee
 where  
  $p=p_F +p_S$  is the total four-momentum of the system, with $p^2=M^2$ 
   ($M$ is the mass of the bound system), while 
     the relative four-momentum is given by 
 $k=\eta_1 p_F-\eta_2 p_S$ (N.B. $\eta_1+\eta_2=1$). We use $\eta_i=1/2$, obtaining  $$p_{F(S)}={p\over
 2}\pm k~~.$$
 The  conjugate BS amplitude is
   obtained analyzing the residue of the 4-leg Green's function
 at the bound pole (assuming for the sake of simplicity, to be only one), and it
 reads
 \be
 \bar \Phi^\pi(k,p,J_z)=\int {d^4 x} ~e^{-ik\cdot x}~\langle \pi, J_z,J;p| 
 T\{\bar \psi(x/2)~\phi^*(-x/2)\}|0\rangle
 \ee 
As it is well-known, the  BS amplitude for a bound state fulfills  the following homogeneous BSE,  where we discard, at the present
stage of our investigation, both self-energy and vertex corrections,
\be
\Phi^\pi(k,p,J_z)=G_0(p/2-k)
S(p/2+k)\int {d^4k^\prime\over (2 \pi)^4}~i{\cal K}^{Ld}(k,k^\prime,p)
~\Phi^\pi(k^\prime,p,J_z)~,\nonu
\label{bse_1} \ee
with
 the relevant  propagators  given by
  \be 
 G_0(q)~=i~{1 \over (q^2 -m^2_S+i\epsilon)}~~, \quad \quad S(q) ~= i~  {\psla q + m_F\over ( q^2 -m^2_F + i \epsilon)}
 \ee

In our calculations in ladder approximation, we adopt the  following momentum-dependent kernels 
for scalar and vector exchanges
\be
i{\cal K}^{Ld}_s(k,k',p)= -i ~\lambda^s_S \lambda^s_F ~{1 \over (k-k')^2
-\mu^2 +i\epsilon}~~,
\ee
and 
\be
i{\cal K}^{Ld}_v(k,k',p)= -i ~\lambda^v_S \lambda^v_F~{(\psla p -\psla k -\psla k')  \over (k-k')^2
-\mu^2 +i\epsilon}
\ee
with $\mu$ the mass of the exchanged boson.

Actually the interaction kernel for the scalar case does not
depend upon the total momentum of the system, while in the vector-exchange case
it does, since the bosonic current is  the sum of the
initial and final momenta. Moreover, it should be pointed out that  the  propagator 
of the exchanged-vector is given in the Feynman gauge.

{ Aim of the present work is to study Eq. \eqref{bse_1} by using both the  NIR 
of the BS
amplitude and the LF-projection technique   in order to  obtain an
eigen-equation formally equivalent to the initial BSE. After that, one can proceed to
numerically solve the  eigenvalue problem and calculate several
quantities and functions through which it is possible to investigate more deeply the inner dynamics. In
particular, in correspondence to the two aforementioned interactions, one can
calculate, e.g., i) the relevant correlation between the binding energy of the system and
the coupling constant, ii) the probability of the valence component, as well as iii)
the longitudinal and transverse LF-momentum distributions.
 The peculiar features of the LF distributions allow one to shed
some light on intriguing effects, that noteworthy
 manifest themselves after carrying out non-trivial dynamical calculations. The proposed  
    physical interpretations, in terms  of   small components of the constituent fermion and its
 polarization, seem to herald new interesting analysis, particularly 
 for the vector
 interaction, where the onset of a scale-invariant regime could appear
 beside a clear evidence of the effect of the helicity conservation, for the
 larger values of the binding energy considered in the present work. 
 
 The paper is organized as follows. In Sect. \ref{sect_BSENIR} the general
 formalism of NIR is introduced and the eigenvalue problem formally equivalent
 to the ladder BSE is worked out; in Sect. \ref{sect_valp} the probability and
 the LF distributions are defined; in Sect. \ref{sect_numres} the numerical
 results are thoroughly presented and discussed. Finally, 
 in Sect. \ref{sect_concl},
 conclusions are drawn and some  interesting perspectives shortly illustrated.} 
\section{BSE and the Nakanishi integral representation}
\label{sect_BSENIR}
In order to solve Eq. \eqref{bse_1}, one proceeds through three main steps 
(for the two-fermion system see, e.g., \cite{CK2010,dFSV1,dFSV2}),
namely:  i) writing down  the  
 most general expression of the
 BS amplitude $\Phi^\pi(k,p,J_z)$ for the system under scrutiny, 
  ii) introducing the 
   NIR and iii) projecting  Eq. \eqref{bse_1} onto the null-plane
   $x^+=x^0+x^3=0$. This series  of operations allows to get the desired solutions
   in Minkowski space.
 
  The fermion-scalar BS amplitude 
  has a Dirac index, and after exploiting Lorentz
invariance,  parity and  the Dirac equation for the whole system, 
it can be
written as follows
\be
\Phi^\pi(k,p,J_z)=\Bigl[O_1(k)~\phi_1(k,p)+ O_2(k)~\phi_2(k,p)\Bigl]~ U(p,J_z)
~~,\label{bsa}\ee
where  $\phi_i$ are unknown scalar functions that depend upon the available
momenta and are   determined by solving the
BSE. The operators $O_i$ act on the spinor $U$ (with normalization $\bar U~U=1$) and one has
\be
O_1(k)= \mathbb{I} ~~,\quad \quad O_2(k)= {\psla k\over M}~~, \quad \quad 
(\psla p - M)~U(p,J_z)=0~~.
\ee
In order to get  the equations fulfilled by the scalar functions $\phi_i(k,p)$, one can 
 multiply both sides of Eq. \eqref{bse_1} by $O_i(k)$, and 
 evaluate  the  following traces, ${\cal N}_{ij}$ and  ${\cal T}_{ij}$,
\be
{\cal N}_{ij}=Tr \left[O_i(k)~O_j(k)~
{(\psla p +M)\over 2M} \right]
\nonu
{\cal T}^{s(v)}_{ij}(k,k',p)=Tr \left[O_i(k)~(\psla p_F +m_F)~\Gamma^{s(v)}~O_j(k')~
{(\psla p +M)\over 2M} \right]~~,
\label{traces}\ee
{ where $\Gamma^s=1$ and  $\Gamma^v=\psla p -\psla k -\psla k'$}. Through this formal elaboration,
one is able to  transform Eq. \eqref{bse_1} into an equivalent 
 coupled system of integral equations for the scalar functions $\phi^{s(v)}_i(k,p)$, viz
\be
\phi^{s(v)}_i(k,p) ={ i \over
( p / 2  -k)^2- m^2_S+i\epsilon  }~
{ i\over
( p / 2  +k) ^2- m^2_F+i\epsilon  } \int { d ^ 4 k' \over (2\pi )^4   }
 \nonu
\times  ~{ (-i {\lambda^{s(v)}_S} \lambda^{s(v)}_F) \over (k-k')^2 -\mu ^2 +i\epsilon  }
~\sum_{j=1,2} {\cal C}^{s(v)}_{ij}(k,k',p)~\phi^{s(v)}_j(k',p)
~~,\label{bse_coup}\ee
with
\be
{\cal C}^{s(v)}_{1j}(k,k',p)=~{M^2\over 2}~
{k^2 {\cal T}^{s(v)}_{1j}  -\spro{k}{p}  {\cal T}^{s(v)}_{2j} \over k^2M^2 -\spro{p}{k}^2} 
\nonu
{\cal C}^{s(v)}_{2j}(k,k',p)=-{M^2\over 2}~{\spro{k}{p}
{\cal T}^{s(v)}_{1j}  - M^2  {\cal T}^{s(v)}_{2j} \over k^2M^2 -\spro{p}{k}^2}
~~.\ee
For the sake of simplicity, in what follows  we drop out the notation
 ${s(v)}$, indicating the type of
interacting kernel one is considering,
but it will be restored when needed. 
The actual expressions of ${\cal C}_{ij}(k,k',p)$, for the scalar and vector 
exchanges are given 
in
 Appendices \ref{scal} and \ref{vect}, respectively.

As in the  case of  a system composed by two scalars \cite{CK2006,CK2006b,FSV2,FSV3,Tomio2016} 
or  by two fermions
\cite{CK2010,dFSV1,dFSV2}), one can introduce the NIR for each
 scalar function $\phi_i(k,p)$ \footnote{Let us remind that the general Dirac structure of a n-leg transition
  amplitude,  with spin dof involved,
 stems from    the combinations of the Dirac structures  in the numerators of each loop.
   This observation leads to 
 the  Dirac structure of the amplitude shown  in Eq. \eqref{bsa}.}
\be
\phi_i(k,p)=\int _{ -\infty  }^{ \infty  }
d\gamma' \int _{ -1 }^{ 1 } dz'\frac {  g_i (\gamma' ,z';\kappa^2) }
{ [{ k }^{ 2 }+z'p\cdot k-\kappa ^2-\gamma' +i\epsilon ]^3}
~~,\label{phi}\ee
where the real functions
 $g_i(\gamma,z;\kappa^2)$ are called  Nakanishi weight functions (NWFs),  
 that depend upon
real variables, and
\be
\kappa^2 =\bar m^2 -{M^2\over 4}~~,
\ee
with $\bar m=(m_F+m_S)/2$.

 In order to complete this first part, we mention that in Appendix \ref{normbsa} the actual
expression of the BS-amplitude normalization, both in terms of the
scalar functions $\phi_i$ and the NWFs  $g_i$, is presented.
\subsection{Determining  the Nakanishi weight functions} 
The appealing motivation for using the expressions in Eq. \eqref{phi} 
 as trial functions for solving the homogeneous BSE,
 is the possibility to make apparent the analytic structure of the BS
amplitude, as  dictated by the analysis performed by Nakanishi in a 
perturbative framework \cite{nak71}. It is important to stress that the 
validity of this procedure  is numerically demonstrated a posteriori, 
i.e. at the end of the {\em formal 
elaboration} we are going to carry out, without further assumptions beyond the
one
shown in Eq.
\eqref{phi}.
 As a matter of fact,
one eventually gets a generalized eigenvalue problem, and  if one finds
   solutions, acceptable from the physical point of
view (i.e. real eigenvalues),
then one can state that the expression  in Eq. \eqref{phi} is flexible 
enough to obtain
actual solutions of the equivalent BSE. 

The last main  step is the so-called LF projection of the BSE, since  it  is based 
 on  the introduction of   LF coordinates $k^\pm=k^0\pm k_z$. As it is well-known, a
  practical advantage in adopting these variables
 is the possibility
  to  split multifold poles in the variable $k^0$ in poles
  for the variables $k^+$ and $k^-$. This simple observation
 (that can be rephrased in a  different formal environment  given in Refs. 
 \cite{CK2006,CK2010}, where the explicitly-covariant LF framework is adopted)
  becomes crucial for obtaining a   substantial simplification of
 the analytical integrations one has to face with in what follows (for the sake
 of comparison see the two-scalar case presented in Ref. \cite{Kusaka}).
The   LF projection of the BSE amounts to integrate  both sides of Eq.
\eqref{bse_coup} on  $k^-$ (see, e.g., \cite{FSV1} and references therein quoted). Notice that such a formal step 
  means to restrict the relative
LF-time to a vanishing value. The main advantage of applying the LF projection
(or the equivalent approach in Refs. \cite{CK2006,CK2010}) is given by the
formally exact reduction of the 4D BSE into an equivalent coupled system for
determining the NWFs $g_i(\gamma,z;\kappa^2)$. However, one should bear in mind that 
the LF
projection of the BS amplitude produces another important outcome, since it
allows
one  to obtain the valence component of 
the interacting state, so that 
 a probabilistic content can be usefully established in the BS approach.

 The LF projection of the scalar functions $\phi_i$, Eq. \eqref{phi}, reads
  (see
 \cite{FSV2,dFSV1,dFSV2} for details)
\be
\int _{ -\infty  }^{ \infty  } \frac { d{ k }^{ - } }{ 2\pi  }\int _{ -\infty  }^{ \infty  }
d\gamma' \int _{ -1 }^{ 1 } dz'\frac {  g_i (\gamma' ,z';\kappa^2) }
{ [{ k }^{ 2 }+z'p\cdot k-\kappa ^2-\gamma' +i\epsilon ]^3}
=\nonu
= { -i \over M } \int _{ -\infty  }^{ \infty  } d\gamma '~\frac {  g_i(\gamma ',z;\kappa^2) }
{  \left[ \gamma '+\gamma + (1-z^2)\kappa^2 +z^2 \bar m^2-i\epsilon  \right]^2   }
~~,\label{Lhs}
\ee
where $\gamma=|{\bf
k}_\perp|^2$,  $z=- 2k^+/M$.

Let us apply  the LF projection also to the rhs of Eq. \eqref{bse_coup}, 
in strict analogy
to the fermionic case \cite{dFSV1,dFSV2}, but with a substantial 
simplification in the
treatment of the LF singularities, generated by the behavior along the arc for
large $k^-$ in the complex plane (see also Ref. \cite{Yan} for the first
discussion of those singularities and the method to fix them). In particular,
 the mentioned 
LF singularities   do not affect the fermion-scalar case, and one can
write the following coupled system
\be
\int _{ \gamma_{min}  }^{ \infty  } d\gamma '\frac {  g_i(\gamma ',z;\kappa^2) }
{  \left[ \gamma '+\gamma + (1-z^2)\kappa^2 +z^2 \bar m^2-i\epsilon  \right]^2   }
=  {\lambda_F~\lambda_S\over 2(4 \pi)^2 }~{1 \over D_0(\gamma,z)}~
 \nonu
\times~\int_0^1dv~v^2
\int_{\gamma_{min}}^\infty d\gamma'\int_{-1}^1 dz'
~\sum_{j=1,2}  ~~g_j(\gamma',z';\kappa^2)
\nonu
\times\left[ {(1+z)^2~B_{ij}( k^-_u)
\theta(z'-z)
\over D_u^2(z',z,m^2_S)}
 +{(1-z)^2~B_{ij}( k^-_d)\theta(z-z')\over D_d^2(z',z,m^2_F)}\right]
~~,\label{nakafin}\ee
where 
\be
D_0(\gamma,z)=\gamma +(1-z^2) \kappa^2+ (\Delta-z\bar m)^2~~,
\ee
and $\gamma_{min}= -2z\bar m|\Delta |+\Delta^2$, 
with
$ 
\Delta=(m_S -m_F)/ 2$. The lower extremum $\gamma_{min}$ is determined in order to avoid
 a free propagation  in the BS amplitude of a bound state,
 i.e. by requiring  the absence of  cuts. The  denominator $D_u$ (for $z'>z$) is
\be
D_u(z',z,m^2_S)=
v(1-v) ~(z'-z)~\left[\gamma -(1-z^2) {M^2\over 4}+m^2_S
\right]
\nonu  +
(1+z)~\left[ v(1-v) \left(\gamma +z^2 {M^2\over 4}\right)+v(\gamma'  +\kappa^2)
+v^2 z'^2 {M^2\over 4}
+(1-v)\mu^2\right]~~,
\ee
Notice that
\be
\lim_{z'\to z}~{D_u(z',z,m^2_S)\over (1+z)}= v(1-v) 
~\gamma + v z^2 {M^2\over 4} +
v(\gamma'  +\kappa^2)
+(1-v)\mu^2~~,
\ee
and therefore, in this limiting case, the factor $(1+z)^2$ in the numerator is
exactly canceled. 

When $z>z'$, the  denominator  is $
D_d(z',z,m^2_F)=D_u(-z',-z,m^2_F)$, and the  same limit for $z'\to z$ is
obtained for $D_d(z',z,m^2_F)/ (1-z)$.

Moreover, in Eq.\eqref{nakafin} one has 
\be
\begin{array} {l l l}B_{11}( k^-_{u(d)})=
c^{(0)}_{11}+c^{(1)}_{11} k^-_{u(d)}~,
& ~~~~~ &
B_{12}( k^-_{u(d)})=
c^{(0)}_{12}+c^{(1)}_{12} k^-_{u(d)}
\\
B_{21}( k^-_{u(d)})=
c^{(0)}_{21}~, 
&~~~~~ &
B_{22}( k^-_{u(d)})=
c^{(0)}_{22}+c^{(1)}_{22} k^-_{u(d)}
\end{array}~~,
\label{Bcoeff}\ee
where the coefficients $c^{(i)}_{jk}$ for scalar and vector exchanges are given in
Appendices \ref{scal} and \ref{vect}, respectively, and 
\be
k^-_u=\frac { M }{ 2 } -{2 ~(\gamma+m^2_S)\over M (1+z)}= 
{~p^-\over 2} - p^-_{S, on}~,
\quad
k^-_d=-\frac { M }{ 2 } +{2~(\gamma+m^2_F) \over M(1-z)}=-{~p^-\over 2} + p^-_{F, on} 
~~.
\label{kud}\ee
It is very important to remind that $k^-_u$ corresponds to have the scalar
constituent on its mass-shell, and hence the fermion is highly virtual, while
for $k^-_d$ the opposite happens.

 The key point is to recognize Eq. \eqref{nakafin} as a generalized eigenvalue
problem. 
The eigenvectors are the  pair of NWFs $\{g_1,g_2\}$, and  
 the corresponding eigenvalues are the product of the coupling constants $\lambda_F \lambda_S$
(or quantities proportional to them, see below).  Once the mass $M$ of the system 
is assigned,  one can proceed
through  standard numerical methods (cf Sect. \ref{sect_numres}).  Indeed, 
the coupled system depends non linearly upon the mass of the 
system, that can be written as
$$M=2\bar m -B$$
where $B$  is the binding energy. The acceptable values of $B/\bar m$ fall in the range
 $
[0,2]$ (see Refs. \cite{Baym,Savkli} for a discussion of the critical behavior of a $\phi^3$
model).

\section{ Probability and  LF-momentum
distributions of the valence component}
\label{sect_valp}
As it is well-known, the BS amplitude does not have  a probabilistic
interpretation, while exploiting a  LF Fock expansion of the interacting state 
\cite{Brod_rev} one can  
retrieve a probabilistic framework, so important and helpful for our physical 
intuition. To accomplish this, one exploits the LF wave functions, i.e. the amplitudes
of the  Fock expansion of the interacting states (in the present case
$~|(1/2)^+\rangle~$). They are invariant under LF boosts, and fulfill the
following normalization constraint (see Appendix \ref{app_vale} for notations and details)
\be
2 (2\pi)^3\sum_n  \int \big[d \xi_i\big] 
\left[d^2 \bmm{\kappa}_{i\perp }\right]
\,\left\vert \psi^{J\pi}_n(\{\xi_{i}\}_n ;\{{\bmm \kappa}_{i \perp }\}_n;\{\sigma_i\}_{n_F} ;J_z)
 \right\vert^2 = 1 ~~,
\label{focknor}\ee
where (i) $ \psi^{J\pi}_n$ are the LF wave functions, i.e. the amplitudes of the Fock
state with $n$ particles (fermions and bosons) and (ii) { $\{\xi_i, 
{\bmm \kappa}_{i \perp }\}$  are the LF-boost invariant kinematical variables
of  the i-th particles \cite{Brod_rev}.}
 The LF wave function with the lowest number of
 constituents, i.e. one fermion and one scalar,
  is the {\em  valence} wave function. From Eq. \eqref{focknor}, it follows that
   the probability to find the valence
  component in the interacting state with $J=1/2$ and given $J_z$ is
  (see also Appendix \ref{app_vale})
\be 
P_{val}={1 \over (2 \pi)^3}~\sum_{\sigma_1}\int_0^1 {d\xi \over 2 ~ \xi (1-\xi)} 
\int
d^2 {\bmm \kappa}_{\perp }~
\left\vert \psi^{J\pi}_{n=2}(\xi  ;{\bmm \kappa}_\perp;\sigma_1 ;J_z)\right|^2
~~.
\label{probv}\ee
After establishing the definitions  within the Fock-expansion framework, it is
compelling to find the relation between the valence component and the 
 BS amplitude, { given its relevance in the application to hadron physics. 
 In view of this, it is useful to recall that the intrinsic
 description of the system contained in the valence component is the one
 living onto the $x^+=0$ hyperplane, as one can straightforwardly realize
 performing  the 4D Fourier transform of the valence component, that has no
 dependence   on $k^-$ (conjugated to the LF-time $x^+$).
 Moreover, one of the external legs of the BS amplitude is a fermion,
  and one should avoid
 any possible kinematical singularity  related to the instantaneous propagation
 (i.e. with $x^+=0$), since the valence component does not have such an analytic
 behavior. This amounts to cut in the fermionic propagator the term
 $\gamma^+/(2k^+)$. The announced relation is (see Appendix \ref{app_vale} for
  more
 details)
 \be
 \psi^{J\pi}_{n=2}(\xi  ;{\bmm \kappa}_{\perp};\sigma_1 ;J_z)
={(1-\xi)}~\sqrt{m_F\over 2}\int {dk^-\over 2\pi} ~
\bar u_\alpha(\tilde q_1,\sigma_1)
\gamma^+_{\alpha \beta} \Phi^\pi_\beta(k,p;J_z)
\label{valBS} 
 \ee
 with $\tilde q_1 \equiv \{ \xi p^+,{\bmm \kappa}_{\perp}\}$ and ${\bf p}_\perp
 =0$. } 
 
 { By introducing in the rhs of Eq. \eqref{valBS} the expression of the BS
 amplitude \eqref{bsa}, one  finds the following components of the 
   valence 
 wave function in terms  of
 the scalar functions $\phi_i(k,p)$, 
  (cf  Eq. \eqref{Lhs})} 
 \be
\tilde\phi_i(\xi,\gamma;\kappa^2)= iM\int_{-\infty}^\infty {dk^-\over 2\pi} \phi_i(k,p)=
 \int _{ -\infty  }^{ \infty  } d\gamma '~\frac {  g_i(\gamma ',z;\kappa^2) }
{  \left[ \gamma '+\gamma + (1-z^2)\kappa^2 +z^2 \bar m^2-i\epsilon  \right]^2   }
~~, \label{valwf}\ee
where \be
\xi={q^+_1\over p^+}= {k^+\over p^+}+{1 \over 2}={1-z\over 2}
\label{xidef}
\ee
Hence,  for the LF valence wave function one writes
\be
\psi^{J\pi}_{n=2}(q^+_1/p^+ ;{\bf q}_{1\perp};\sigma_1 ;J_z)=
-i~(1-\xi)~\sqrt{\xi\over 2M}
\nonu \times ~\left\{ \delta_{\sigma_1,J_z}\left[
\tilde\phi_1(\xi,\gamma;\kappa^2)
-{z\over 2} ~\tilde\phi_2(\xi,\gamma;\kappa^2)\right]
- \delta_{-\sigma_1,J_z}  ~2J_z~
{k_x +  i 2J_z k_y\over M} ~\tilde\phi_2(\xi,\gamma;\kappa^2)
\label{psi2spin}\right\} ~~.
 \ee
Notably, in Eq. \eqref{psi2spin}  the two contributions stemming from 
the configurations with  the spins of the constituent and the system   
 {\em aligned}
or {\em anti-aligned}  are well identified. 
Finally combining Eqs. \eqref{probv} and \eqref{psi2spin}, one writes
\be
P_{val}=P^A_{val}+P^{noA}_{val}
\ee
where the  probabilities of the aligned and anti-aligned configurations are
given by
\be
P^A_{val}= {1 \over 32 M\pi^2}~\int_0^1 {d\xi ~(1-\xi)} 
\int_0^\infty d{\gamma} ~\left[ \tilde\phi_1(\xi,\gamma;\kappa^2)-
{z\over 2}\tilde\phi_2(\xi,\gamma;\kappa^2) 
\right]^2
\nonu
P^{noA}_{val}={1 \over 32 M\pi^2}~\int_0^1 {d\xi ~(1-\xi)} 
\int_0^\infty d{\gamma} ~
{\gamma\over M^2}~
\tilde\phi^2_2(\xi,\gamma;\kappa^2)~~.
\label{probv1}\ee
Another set of quantities quite relevant for understanding the dynamics in the valence
component, and consequently interesting from the experimental point of view, is given
by 
the LF valence distributions, that  describe  i) the probability distribution to find a
constituent with a given longitudinal fraction $\xi$ and ii) the 
probability distribution to 
find a constituent with transverse momentum $\gamma=|{\bf k}_\perp|^2$.
They are defined for the  fermionic constituent as follows
\be
\phi^F(\xi)={1 \over 32 M\pi^2}~ ~(1-\xi) 
\int_0^\infty d{\gamma} ~\left[ \left(\tilde\phi_1(\xi,\gamma;\kappa^2)-
{z\over 2}\tilde\phi_2(\xi,\gamma;\kappa^2)\right)^2 +
{\gamma\over M^2}~
\tilde\phi^2_2(\xi,\gamma;\kappa^2)
\right]
\label{longdis}\\ &&
{\cal P}^F(\gamma)={1 \over 32 M\pi^2}~\int_0^1 {d\xi ~(1-\xi)} 
 ~\left[ \left(\tilde\phi_1(\xi,\gamma;\kappa^2)-
{z\over 2}\tilde\phi_2(\xi,\gamma;\kappa^2)\right)^2 +
{\gamma\over M^2}~
\tilde\phi^2_2(\xi,\gamma;\kappa^2)
\right]
\label{trandis}
\ee
and  are normalized to $P_{val}$. One can easily recognize the two contributions: the
aligned and the anti-aligned ones.

\section{Numerical Results}
\label{sect_numres}
The numerical method for solving  the coupled  system in Eq. 
\eqref{nakafin}
strictly follows the one already adopted for the two-fermion case
\cite{dFSV1,dFSV2}. Basically, one   expands the NWFs
 on an  orthonormal basis given by the Cartesian 
product of Laguerre polynomials and Gegenbauer ones, in order to take care of the
dependence  
upon $\gamma$ and $z$, respectively. Unfortunately,
in the case of the fermion-boson system one cannot exploit the symmetry under the 
exchange of the two constituents (i.e. $z\to -z$) for constraining the symmetry of the BS
amplitude, and in turn  of  the NWFs (particularly the odd or even
dependence upon $z$).
Hence,  the adopted orthonormal basis   contains  both symmetric 
and antisymmetric Gegenbauer polynomials.
The expansion of the $g_i$ reads 
\be
g_{i}(\gamma, z;\kappa^2) = \sum_{m=0}^{\infty} \sum_{n=0}^{\infty}
A_{mn}^{i}(\kappa^2)~G^{\nu_i}_{m}(z) \, \mathcal{J}_{n}(\gamma) ~,
\label{exp}\ee
where i) $A_{mn}^{i}$ are suitable coefficients to be determined by solving the
generalized eigen-problem given by the coupled-system \eqref{nakafin}, ii) 
$G^{\nu_i}_{m}(z)$, are related to the Gegenbauer polynomials 
$C_{m}^{\nu_i}(z)$, while  iii) $\mathcal{J}_{n}(\gamma)$  are given in terms of 
 the
 Laguerre polynomials $L_{n}(a\gamma)$. In particular, one has
\be
 G_{n}^{\nu} (z)  =  (1-z^2)^q~~ \Gamma(\nu) 
 \sqrt{\frac{n! (n + \nu)}{ 2^{1-2\nu} \, \pi\, \Gamma(n + 2\nu)}}
~ C_{n}^{\nu}(z) ~,\nonu
\mathcal{J}_{n}(\gamma)  =  \sqrt{a} \, L_{n}(a\gamma) e^{-a\gamma/2} ~.
\label{basis}\ee
where  $ q=(2\nu-1)/4$ has been taken equal to $1$ for $g_1$ and $3$ for $g_2$,
respectively.  The parameter  $a$ governs the fall-off of the NWFs for large
$\gamma$ and   it turns out that when $M$  becomes smaller and
smaller,  
decreasing its value is helpful  from the numerical point of view.
 Indeed, for
smaller values of the mass, the system is more compact, and therefore the  kinetic energy
increases, emphasizing the relevance of the tail in $\gamma$ of the NWFs.  In order to speed up the convergence of the integration on $\gamma$, this
variable has been rescaled by a factor $a_0$, i.e. $\gamma\to \gamma/a_0$. 
{ In the actual
calculation, $a_0$ has been chosen equal to $12$, while $a=6$.}

Two general observations are in order: i) after introducing the above expansion and
the proper projection,  the lhs of \eqref{nakafin} reduces to   a symmetric
real matrix applied to a vector containing the coefficients of the expansion, 
while the rhs contains  a non symmetric matrix, ii) the eigenvalues 
can be real
or complex conjugated. In conclusion, one symbolically  writes
\be
 L(M)~{\bmm v}= \alpha ~R(M)~{\bmm v}
 \ee
 where the non linear dependence upon the mass of the system, $M$, is present in both
 sides. The search of the eigenvalues, i.e. the coupling constant compatible with
 the assigned mass $M$, proceeds by looking for the lowest real eigenvalue,
  { which corresponds to the shallowest well able to support a bound system
  with mass $M$, if one exploits a physical intuition based on a 
  simple, non-relativistic  instance. The smallest eigenvalue} is 
 obtained 
  by using first a low number of basis functions in  the expansion in Eq.
 \eqref{exp},  and then  checking
 the stability of the result by increasing the basis. 
  To determine the
 eigenvectors, i.e. the coefficients in the expansion \eqref{exp}, { requests 
   more 
 care. For
  getting } more
 stable results, following Ref. \cite{CK2006}, a small quantity $\epsilon
 =10^{-8}$  has been added to the diagonal elements of the lhs matrix, and
 furthermore both
 sides of Eq. \eqref{nakafin} have been multiplied by a factor $(1-z^2)^p$,  
 that helps to
 enforce the constraint at the extrema of the variable $z$. The  used values
 for the exponent $p$ is $1$ for the scalar exchange and $2$ for the vector one.
 Finally,   the following numerical results correspond to the maximal { values for $m$ and $n$ in Eq. \eqref{exp}
 $N_{Lag}=N_{Geg}=56$.}

\subsection{Scalar interaction}

In order to get rid of the
dimensional dependence on the mass, the coupling constant for the scalar exchange is defined as
follows 
\be\alpha^{S}=
{\lambda^s_F~\lambda^s_S\over 8 \pi m_S}
\label{alphas}~~.\ee
 Notice that the factor
$8\pi$ is twice  the familiar $4\pi$, { in order to match the non-relativistic
reduction of the Born term in the fermion-scalar scattering, properly taking
into account the different relativistic normalization  
of fermionic and bosonic states.}
Table \ref{tab1} shows the  
 coupling constants for the case of a scalar exchange and equal-mass 
  constituents, i.e. $m_F=m_S$. In particular, the presented  results 
  correspond to 
 binding energies in unit
 mass
$B/\bar m\in[0.1,1]$ and two  values of the exchanged-boson  mass
$\mu/\bar m=0.15,~0.50$.  For the sake of  
comparison, in Table \ref{tab1},    it is also shown 
the analogous set of results evaluated after Wick-rotating the relative
variables $k_0$ and  $k'_0$ in Eq. \eqref{bse_coup}, namely without 
inserting LF
coordinates, but keeping the standard ones and changing both $k_0 \to i k_0$ and
$k^\prime_0 \to i k^\prime_0$. 
It is worth noticing that  the values of the dimensionless coupling constant for the scalar
exchange are larger than the ones shown in Table \ref{tab2}, corresponding
to the vector exchange. Such a difference can be seen also for the two-fermion
case \cite{dFSV2}, and it can be ascribed to the repulsion generated by
the small component of the fermion spinor when a scalar vertex is involved.
As a matter of fact, the scalar
 interaction meets  a  difficulty to bind a system, when it becomes more and
 more compact (i.e. $B$ increases). 
 {In order to identify the source of the repulsion that opposes  the binding,
 one should analyze the low $M$ behavior of the coefficients 
 $B_{ij}(k^-_{u(d)})$
 given in Eq.
 \eqref{Bcoeff}. After inserting   the coefficients shown
  in  Eq. \eqref{Ccoeffs} and 
  $k^-_{u(d)}$ from Eq. \eqref{kud}, one gets that only   
  $B_{12}(k^-_u)$ is always
  negative for $m_S>M/2$ and it becomes larger and larger (in modulus) for $M\to 0$ (strong coupling
  limit), viz
  \be
  B_{12}(k^-_u)=-\frac { z'v }{ 2 } \left( \frac { M }{ 2 } +{ m }_{ F  } \right) 
 -{\left( 1-v \right)\over M (1+z)}~\left[2\gamma+(1-z) \left(m^2_S-{M^2\over
 4}\right)+ z^2 {M^2\over 2}\right]
  \ee
   The numerical checks, obtained after omitting this 
  negative term, 
  show an almost  $40\%$ reduction of the coupling $\alpha^S$,  for
  $\mu/\bar m=0.5 $  and $B/\bar m=1$, with $P_{val}= 0.82$, and the calculations
  can be even extended to  $B/\bar m>1$ without any large increase of the
  coupling constant, as well as a smoother growing of  $P_{val}$. 
  The physical source of the repulsion} can be heuristically understood once 
 we recall that    the  fermion-scalar vertex, when  initial and final fermions are  on-mass shell, 
 contains the scalar density  $\bar u~u$,
    and 
 the Dirac matrix $\gamma^0$  generates a minus
 sign in front of   
 the contribution produced by  the small components. Hence  a repulsion
 is produced. Moreover, one should expect that the
 repulsive effect of the small components of the fermion spinor is driven by the
 kinetic energy, since the scalar density is written in terms of the 
   large  Dirac component, $f$, and the small  one, $g$,  as follows:
 $\rho_s=\bar u u\sim |f({\bf k})|^2-|g({\bf k})|^2\sim \rho_v-  2
 |{\bf k} ~f({\bf k})|^2/(E+m)^2$, with the vector density  given by
  $\rho_v=u^\dagger u\sim |f({\bf k})|^2+|g({\bf k})|^2$.

 \begin{table}
 
 \caption{Scalar coupling $\alpha^S$, Eq. \eqref{alphas}, for
 $m_F=m_S$ and $\mu/\bar m=0.15,~0.50$ 
 (with $\bar m=(m_S+m_F)/2$). 
 First column: the binding energy in unit mass of $\bar m$, i.e. $B/\bar m$.
 Second and fourth columns: coupling constants $\alpha^S_M$, obtained 
 by solving the BSE \eqref{bse_1} in Minkowski space through Eq. 
 \eqref{nakafin}. 
Third and fifth columns: Wick-rotated results, $\alpha^S_{WR}$ (see text). }
\label{tab1}
 \begin{center}
 \begin{tabular}{|l|r|r||r|r|}
 \hline
 $B/\bar m$ 
 & $\alpha^S_{M}(0.15)$ & $\alpha^S_{WR}(0.15)$ &
 $\alpha^S_{M}(0.50)$ &$\alpha^S_{WR}(0.50)$\\
 \hline
 0.10& 1.506 &1.506  & 2.656& 2.656\\
 0.20& 2.297 &2.297  & 3.624 &3.624  \\
 0.30& 3.047 &3.047   & 4.535 &4.535  \\
 0.40& 3.796 & 3.796  &5.451 & 5.451 \\
 0.50& 4.568 & 4.568  &6.404 & 6.404 \\
 0.80& 7.239 &7.239  &9.879 &9.879   \\
 1.00& 9.778      &9.778  & 13.738  & 13.738  \\
 \hline
 \end{tabular}
 \end{center}
 \end{table}
 
For illustrative purpose, in Fig. \ref{fig2}, the 
NWFs
 for the scalar exchange with 
 $\mu/\bar m=0.15$ 
  and equal-mass
 constituents, are presented as a function of i) $\gamma$ and fixed $z=0$ and ii)
  $z$ and fixed $\gamma=0$, respectively. Interestingly, the difference between
  $g_1$  and $g_2$ increases for large binding energies.
   Indeed,  { such an effect could be related to the weight in front of 
   $\phi_2$, i.e. the factor $\psla k/M$ (cf   
   Eq. \eqref{bsa}). As a matter of fact, for increasing $B/\bar m$ the average
   size of the system decreases and 
   large values of the kinetic energy (related to the relative momentum $k$)      
    become more and more likely,
and in order to avoid a blowing  contribution from
  the second term in Eq. \eqref{bsa}, the amplitude 
  $\phi_2$ should decrease. The same happens for larger values of $\mu/\bar m$,
  since  the system becomes more compact, given the 
  shrinking of the
  range of the interaction.} It is worth mentioning that the NWFs can  have
  wild oscillatory behaviors, that fade out in a smooth pattern of the LF
  distributions, discussed in what follows, given the filtering role played
   by the integral kernel 
  in Eq. \eqref{valwf}. Such an effect is  well-known in the
  analysis of signals where the generalized Stieltjes transforms are commonly
  adopted (see, e.g. Ref. \cite{esignal}).
   \begin{figure}
 \includegraphics[width=8cm]{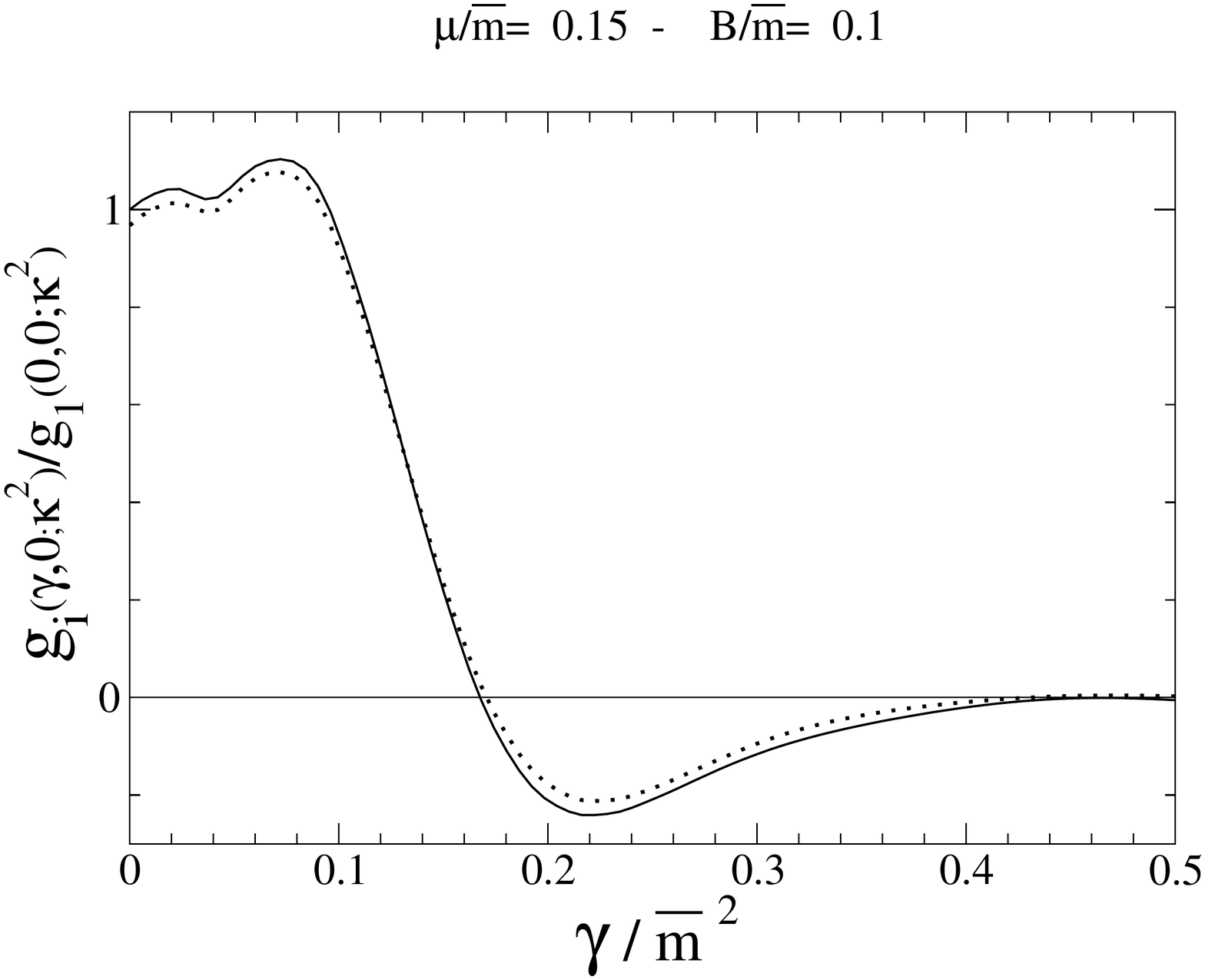}\includegraphics[width=8cm]{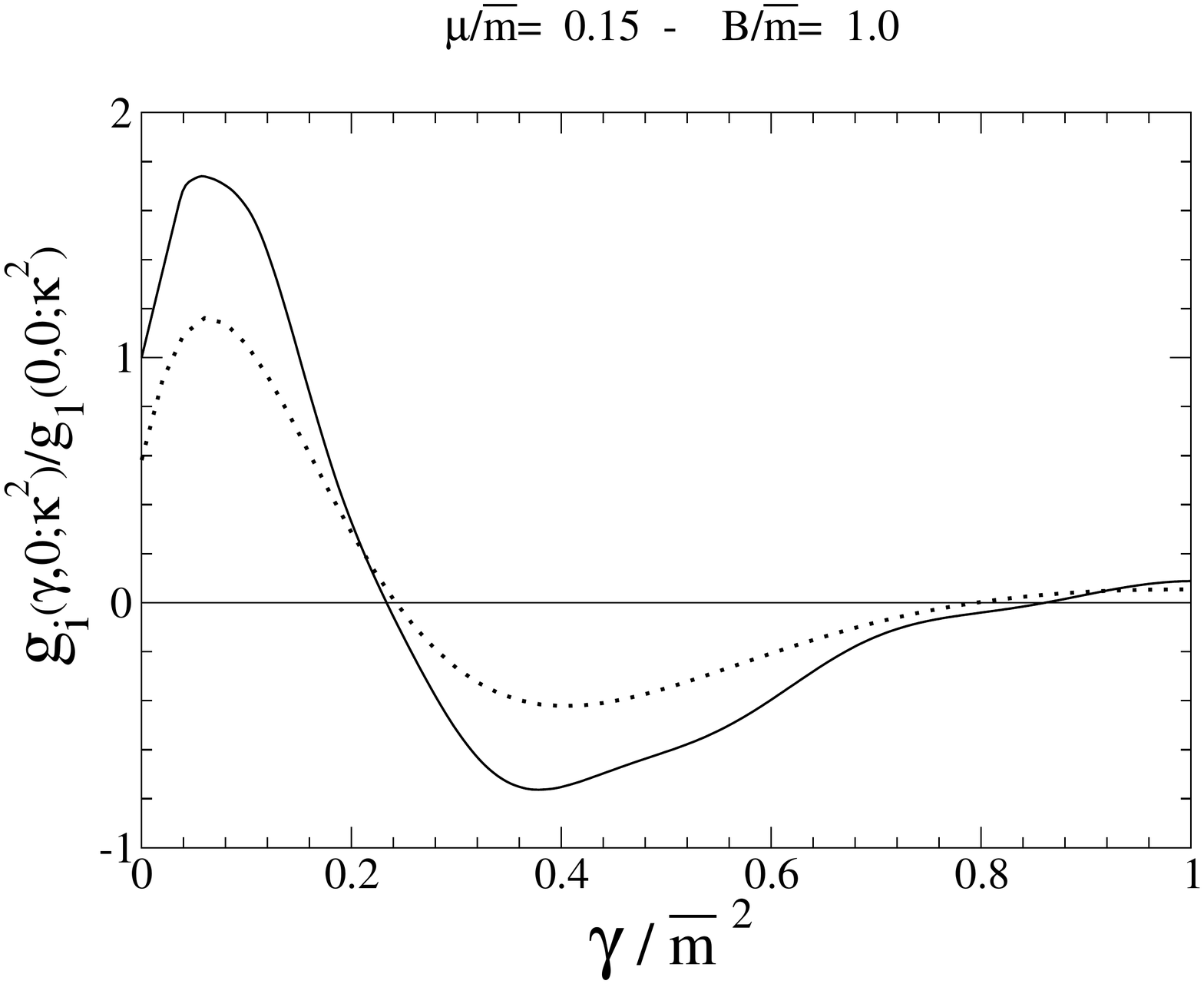}

\includegraphics[width=8.5cm]{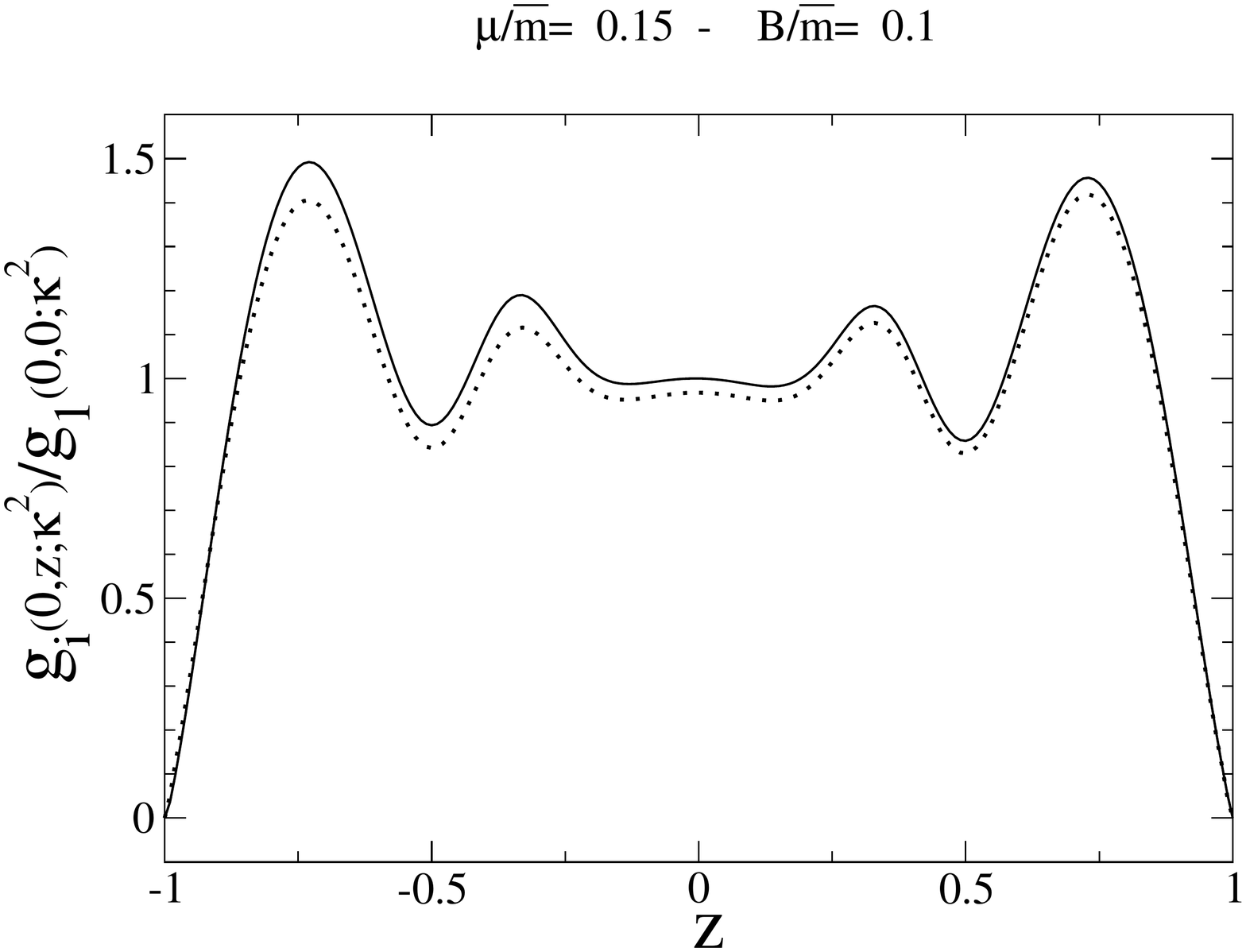}\includegraphics[width=8.5cm]{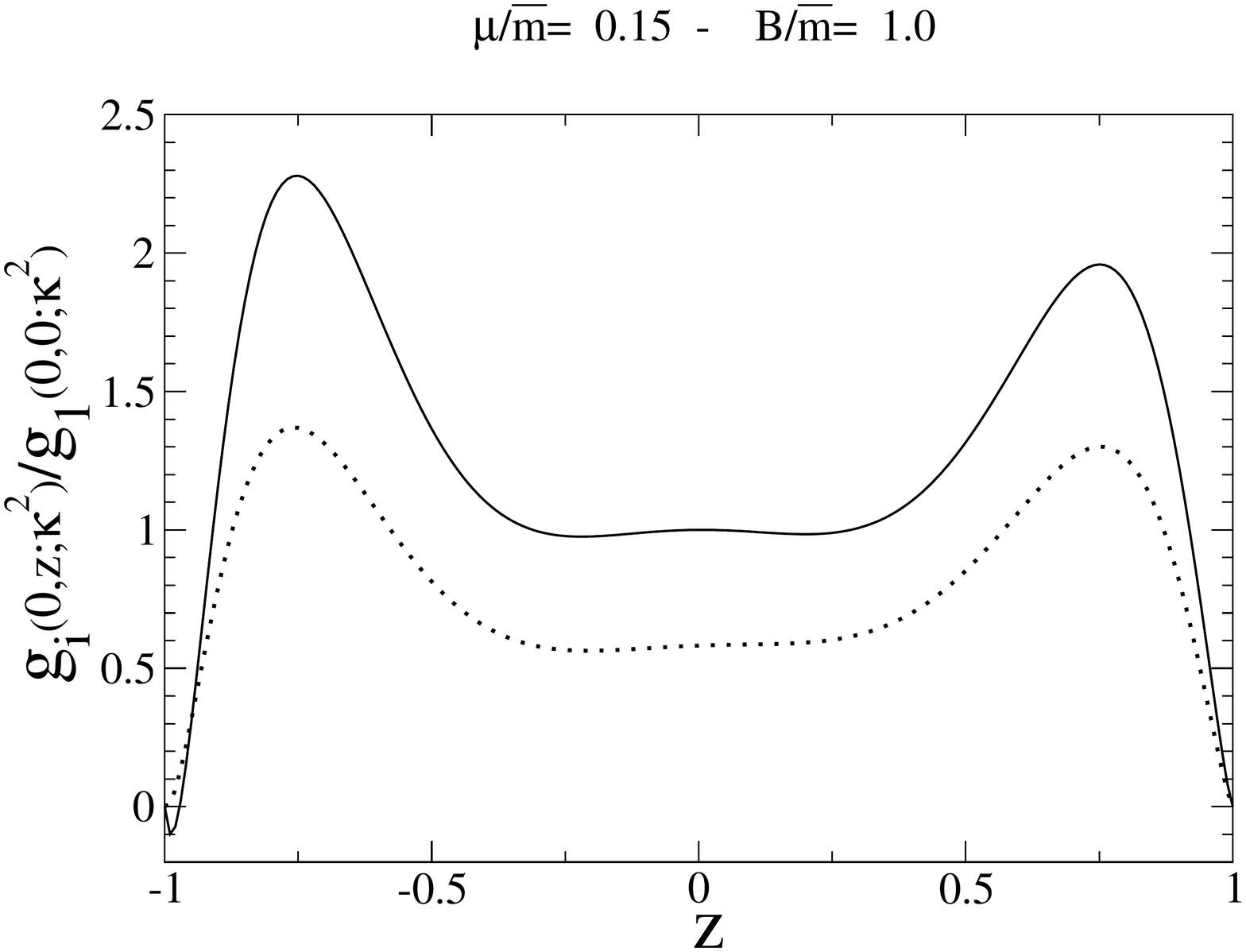}
 \caption{The Nakanishi weight functions $g_i$  for an equal-mass $(1/2)^+$ system, 
 and the interaction mediated
 by a scalar boson with $\mu/\bar m=0.15$. Left side:  the calculations 
 with $B/\bar m =0.1$.  Right side: calculations with $B/\bar
 m=1$. Solid line : $g_1$. Dotted line: $g_2$. Upper panels: $g_i$ vs  $\gamma$, and
 $z=0$. 
   Lower panels:   $g_i$ vs  $z$, and $\gamma=0$.
   Notice that the normalization factor is the same for both NWFs, namely 
   $g_1(0,0;\kappa^2)$.
 }
 \label{fig2}
 \end{figure}
 
 The relevance of the valence component in the Fock expansion of the interacting
state  is illustrated  in Table \ref{tab1p},
 where the
valence probabilities, defined in Eq. \eqref{probv}, are shown together with 
 the two
  contributions from the possible 
configurations of   the spin of the system and the spin of the constituent,
namely
 i)   aligned 
or ii) anti-aligned. In the second case, one must have    
a component of the orbital angular momentum ${ L}=1$ in order to get 
a third component of the
total momentum equal to $J_z=\pm 1/2$. The valence probabilities start 
to behave
in an unexpected way when the binding increases. This seems to indicate
that the repulsion we mentioned above  damps the coupling of the valence state with the higher Fock-components,
and consequently the valence probability increases. { Indeed, the large kinetic energy needed to allow a compact
system (the size is related to the inverse of the binding energy) is more efficiently shared on a two-constituent
Fock state than on  multi-particle ones.}

 Further insights can be gained
 from the  analysis the LF-momentum
distributions, presented in what follows.
 \begin{table}
 \caption{Valence probabilities (see Eq. \eqref{probv}), for the scalar exchange
 with
 $m_F=m_S$ and $\mu/\bar m=0.15,~0.50$. The two contributions, $P^{noA}_{val}$ and
  $P^{A}_{val}$, corresponding to 
 the configurations
 where the spin of the constituent is anti-aligned or aligned 
 to the spin of the system, are also shown. 
 First column: the binding energy in unit mass of $\bar m$.}
\label{tab1p}
 \begin{center}
 \begin{tabular}{|cc||c|}
 \hline  \phantom{$B/\bar m$}&\hspace{0.0cm}$\mu/\bar m=0.15$\hspace{0.55cm} & 
 \hspace{0.25cm}$\mu/\bar m=0.50$\hspace{0.30cm} \\
 \hline 
 \end{tabular}
 
 \begin{tabular}{|l|r|r|r||r|r|r|}
 \hline
 $B/\bar m$ 
 & $P_{val}$ & $P^{noA}_{val}$ & $P^{A}_{val}$
 & $P_{val}$& $P^{noA}_{val}$ & $P^{A}_{val}$ \\
 \hline
 0.10& 0.81 &0.02 &0.79 & 0.88 &0.03 &0.85\\
 0.20& 0.77  & 0.03&0.74&0.85  &0.05 &0.80\\
 0.30& 0.76  &0.05 &0.71& 0.84 &0.07 &0.77\\
 0.40& 0.75  &0.06 &0.69& 0.83 & 0.09&0.74\\
 0.50& 0.76  & 0.07&0.69& 0.83 &0.11 &0.72\\
 0.80& 0.81 &0.13 &0.68 & 0.88 &0.18 &0.70\\
 1.00& 0.90  &0.19 &0.71& 0.98 &0.25 &0.73\\
 \hline
 \end{tabular}
 \end{center}
 \end{table}

In Figs. \ref{fig4} and \ref{fig5}, the longitudinal  and transverse 
 LF distributions (cf Eqs. \eqref{longdis} and \eqref{trandis}),   
are shown for the fermion in the valence
 component, with
  $\mu/\bar m=0.15,~0.50$, and three-values of the binding energy 
  $B/\bar m= 0.1, ~0.5,~ 1$.
  The longitudinal distribution shows a peak around $\xi=0.5$ for $B/\bar
  m=0.1$,  that broadens and sizably reduces its height 
  (slightly shifting towards lower values of $\xi$), when  the
  binding energy and/or the mass of the exchanged boson increase. 
  { The behavior substantially follows the 
   increasing pattern of the
  anti-aligned probability (that involves the orbital $L=1$
  component).  This observation suggests that the fermion, to be considered almost massless
   for large kinetic energy, tends toward 
  a positive helicity. Indeed, the increasing of the tail  at $\xi \ge 0.8$ (for $B/\bar m>1$  a sizable
  bump appears)
  is given  by the aligned configuration, and the large values of $\xi$
   mean a Cartesian 3-momentum
  aligned along the positive $z$-axis. Differently,  the anti-aligned 
  configuration dominates
  the probability at  small values of $\xi$ (i.e. a Cartesian 
  3-momentum along the negative direction of the $z$-axis), and again one
  recovers a preferred positive helicity.
  } 
  Notice that 
  also the transverse distribution  follows a pattern correlated to the
  growing  of the average kinetic energy, as it happens for bigger values of  $B/\bar m$ 
 and/or the exchanged-boson mass.
  
   As a final remark, we mention that the values of the
 average { $<\xi>$ follows a slightly decreasing  pattern  from 
 $0.5$ for  $B/\bar m=0.1$ to $0.41$  at
 $B/\bar m=1$}, almost irrespective of the value of $\mu/\bar m$, while    for 
 $<\gamma/\bar m^2>$   one goes from  $ 0.09$ 
 to   $0.47$ with $\mu/\bar m=0.15$, and  from 
 $0.15$ to   $0.64$ with  $\mu/\bar m=0.50$.
  
\begin{figure}
 \includegraphics[width=9.cm]{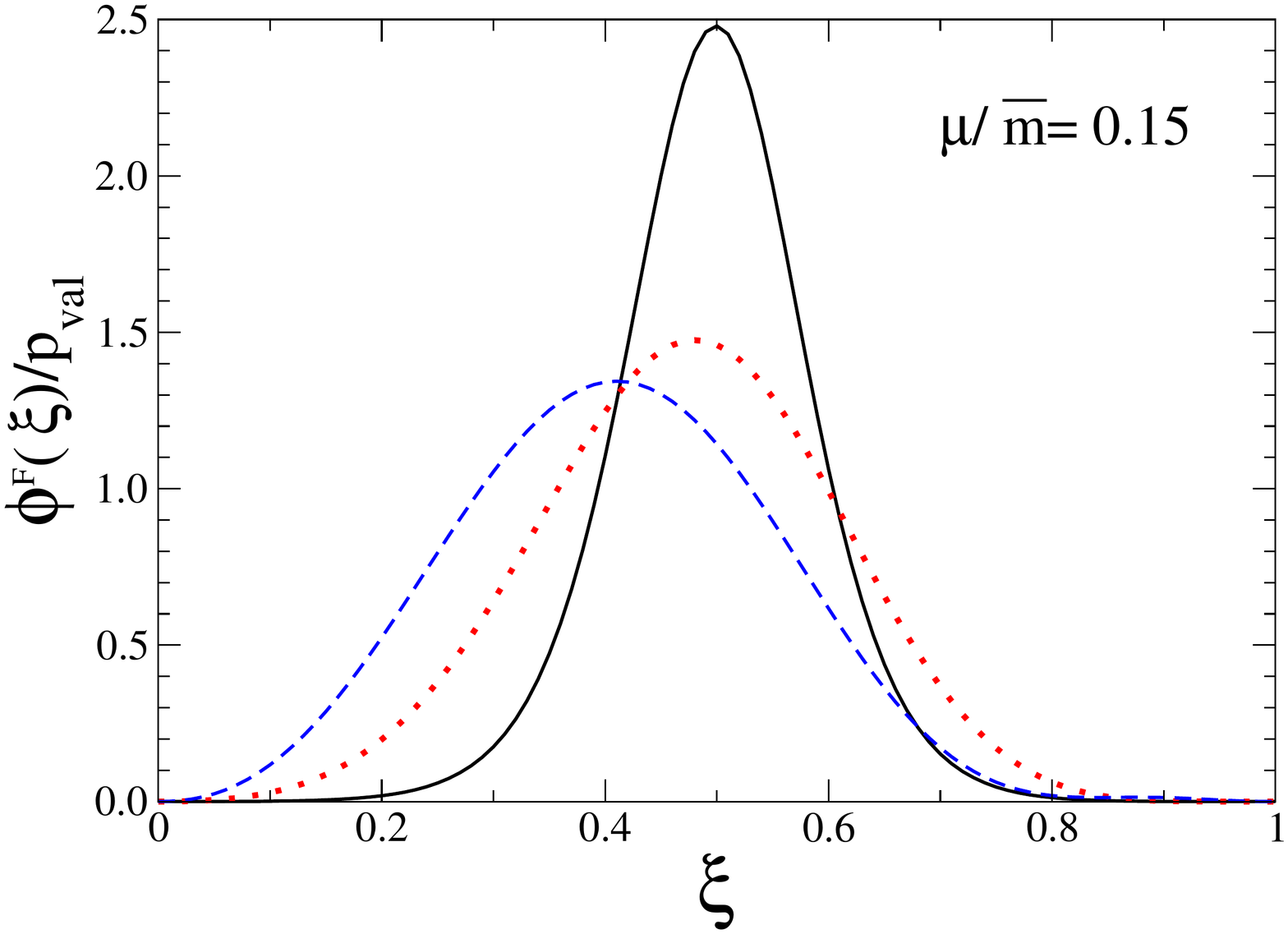} \hspace{-1.8 cm}
 \includegraphics[width=9.cm]{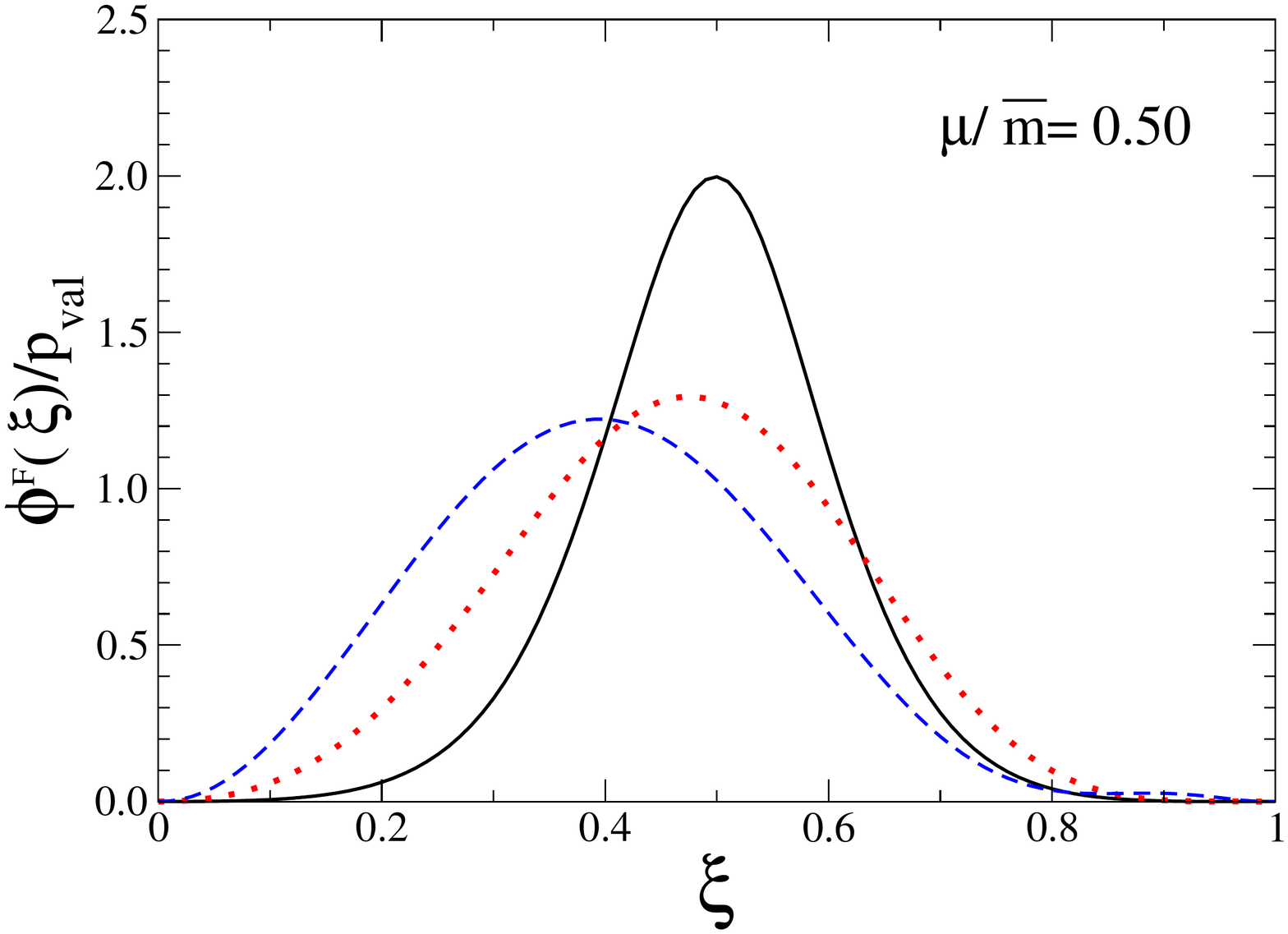}
 \caption{(Color online). Longitudinal LF distributions for  a fermion in the valence
 component for $\mu/\bar m=0.15$ (left panel) and for $\mu/\bar m=0.50$ (right
 panel), in the case of a scalar exchange. Solid line: $B/\bar m=0.1$. Dotted red line: $B/\bar m=0.5$.
  Dashed blue line:
 $B/\bar m=1.0$.}
 \label{fig4}
 \end{figure}
 \begin{figure}
 \includegraphics[width=8.7cm]{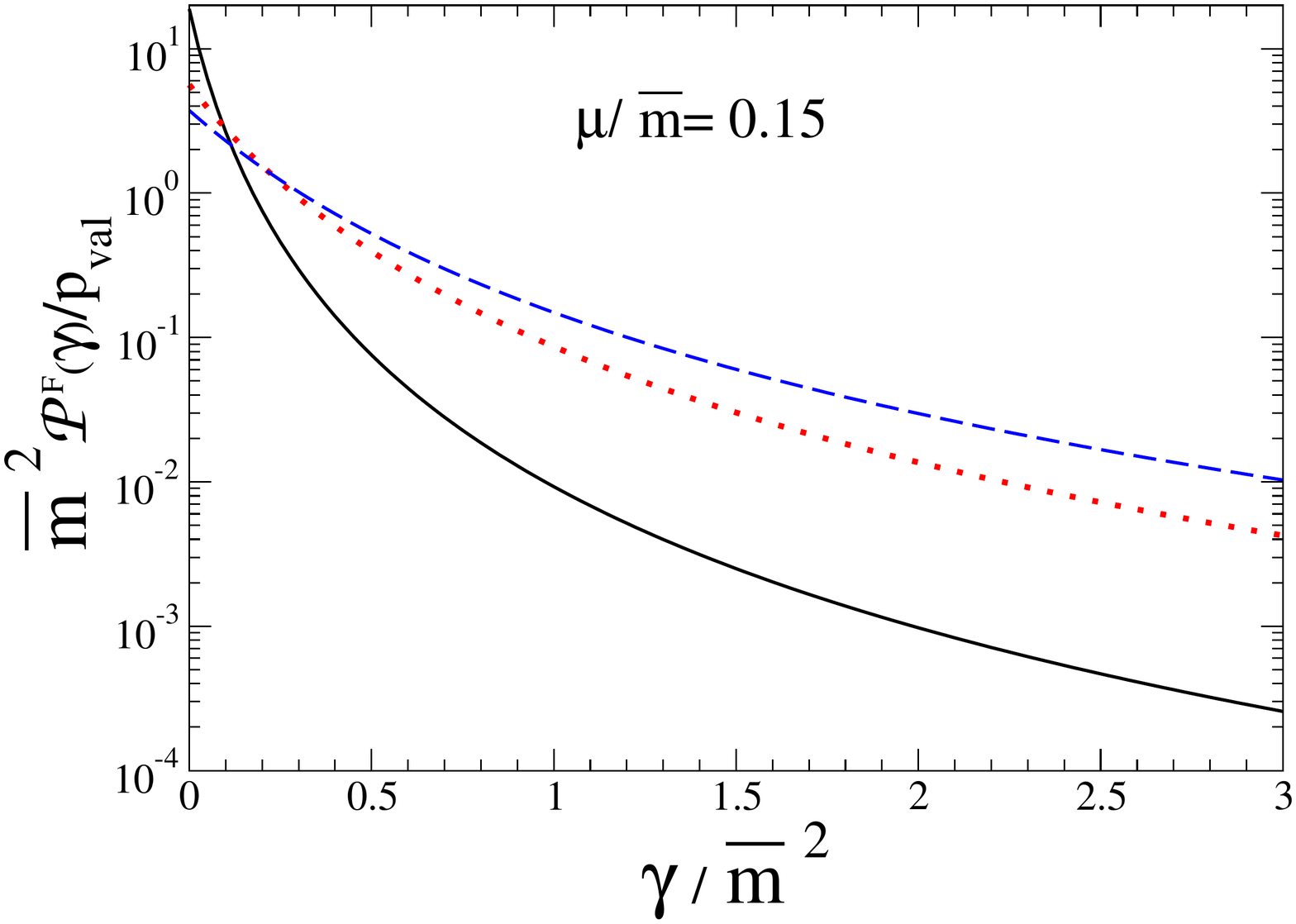}\hspace{-1.3cm}
 \includegraphics[width=8.7cm]{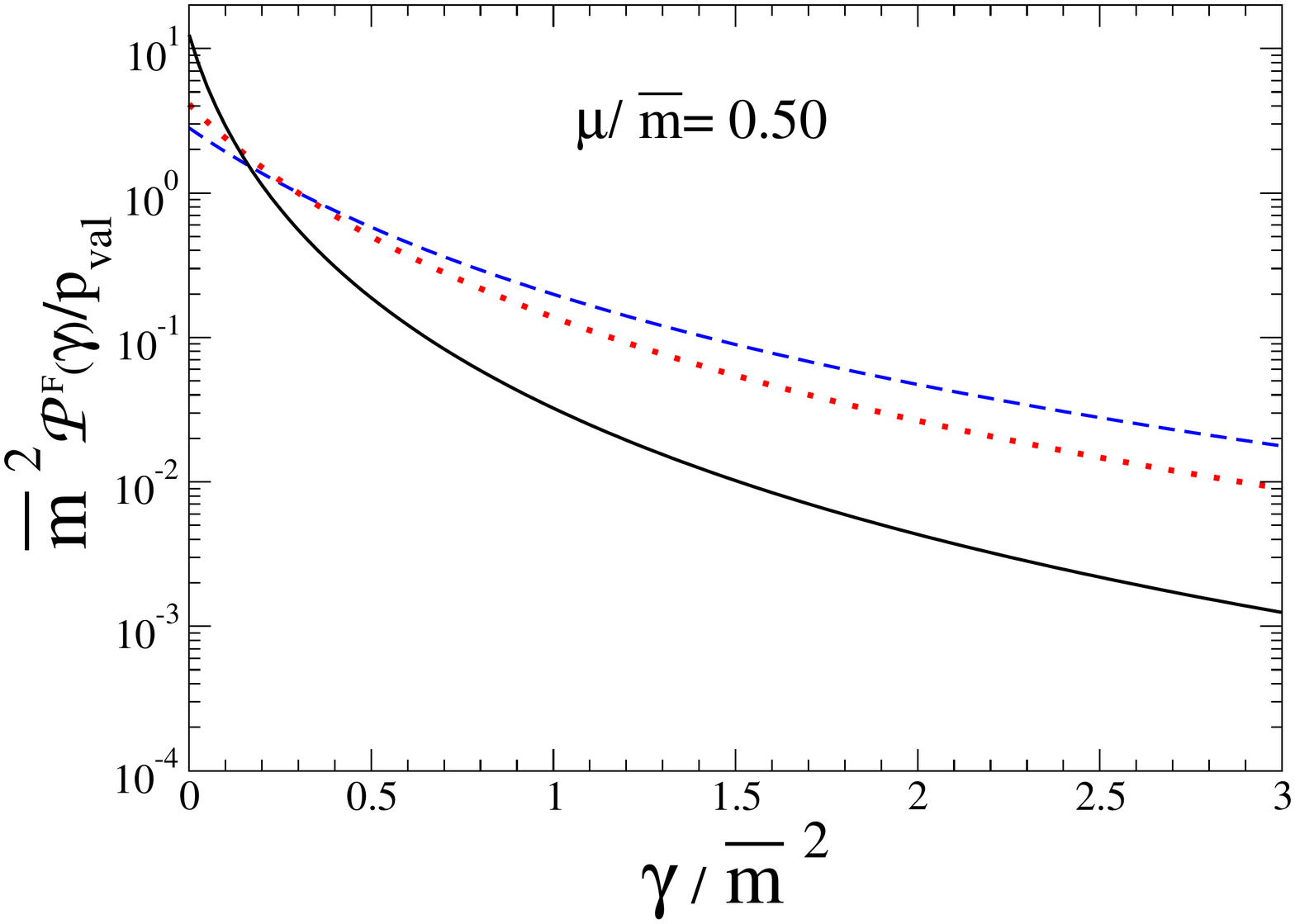}
 \caption{(Color online). Transverse LF distributions for  a fermion in the valence
 component for $\mu/\bar m=0.15$ (left panel) and for $\mu/\bar m=0.50$ (right
 panel), in the case of a scalar exchange. Solid line: $B/\bar m=0.1$. Dotted red line: $B/\bar m=0.5$.
  Dashed blue line:
 $B/\bar m=1.0$.}
 \label{fig5}
 \end{figure}
 
 \subsection{Vector interaction}
 
  For the vector exchange case,  the coupling constant  is defined as
 \be \alpha^{V}=
{\lambda^v_F~\lambda^v_S\over 8 \pi }\label {alphav}\ee
and it  does not contain  any mass in the denominator, given the dimensionless nature of the vertex
 constants in the interaction Lagrangian.  {Being dimensionless 
 the vertex constants, the { BSEs
both in  Euclidean  and in Minkowski spaces, as well as the system of
 integral equations for the 
NWF, have the property to be invariant under a scale transformation
in the ultraviolet region. Such a symmetry imposes a maximum value for the coupling constant, 
beyond  which the invariance 
is broken. One encounters  a similar situation in the fermion-fermion bound state problem in the ladder approximation
 both in Euclidean \cite{Dork} and in Minkowski space \cite{CK2010}. 
 Here, we adopt a conservative point of view and  
 present calculations for moderate bindings, leaving the detailed study of the scale invariance
  breaking, that should establish 
 at  larger bindings, for a future work \cite{scaleinv}. 
  Our results in Minkowski space, shown in  Table \ref{tab2} up to 
  $B/\bar m=0.5$,  nicely agree 
   with the Wick-rotated calculations, analogously to what happens 
   for  the scalar-exchange case.
 }
  
\begin{table}

 \caption{Vector coupling $\alpha^V$, Eq. \eqref{alphav}, for
 $m_F=m_S$ and $\mu/\bar m=0,~0.15,~0.50$. 
 First column: the binding energy in unit mass of $\bar m$, i.e. $B/\bar m$.
 Second,  fourth and sixth column: coupling constants $\alpha^V_M$, obtained 
 by solving the BSE \eqref{bse_1} in Minkowski space through Eq. \eqref{nakafin}. 
Third,  fifth and seventh column: Wick-rotated results, $\alpha^V_{WR}$,   with a
numerical uncertainty for $B/\bar m=0.5$ due to some instabilities in 
the Gaussian quadrature adopted. 
}\label{tab2}

 \begin{center}
 \begin{tabular}{|l|r|r||r|r||r|r|}
 \hline
 $B/\bar m$& $\alpha^V_{M}(0)$ &$\alpha^V_{WR}(0)$ 
 & $\alpha^V_{M}(0.15)$ &$\alpha^V_{WR}(0.15)$ &  
 $\alpha^V_{M}(0.50)$ &$\alpha^V_{WR}(0.50)$\\
 \hline
 0.10& 0.513 & 0.513  &0.608     &0.609   & 0.849  & 0.854\\
 0.20& 0.758 & 0.761  &0.823     &0.823   & 1.009   & 1.015  \\
 0.30& 0.936 & 0.938  &0.979     & 0.978  & 1.127  & 1.129  \\
 0.40& 1.074 & 1.074  &1.107     & 1.097  & 1.225  & 1.216 \\
 0.50& 1.189 & 1.18 $\pm$ .03  &1.214     & 1.19 $\pm$ .03  & 1.311  
 & 1.28 $\pm$ .04  \\
 \hline
 \end{tabular}
 \end{center}
 \end{table}

 In Table \ref{tab2p}, the valence probabilities are shown for the vector exchange. In the range of $B/\bar m$ we have
 investigated, as dictated by the onset of a scale-invariant regime, 
 they smoothly decrease.
 \begin{table} 
 \caption{The same as in Table  \ref{tab1p}, but for the vector exchange.}
\label{tab2p}
 \begin{center}
 \begin{tabular}{|cc||c||c|}
 \hline  \phantom{$B/\bar m$}&\hspace{0.1cm}$\mu/\bar m=0.0$\hspace{0.67cm}
 &\hspace{0.25cm}$\mu/\bar m=0.15$\hspace{0.30cm} & 
 \hspace{0.25cm}$\mu/\bar m=0.50$\hspace{0.30cm} \\
 \hline 
 \end{tabular}
 
 \begin{tabular}{|l|r|r|r||r|r|r||r|r|r|}
 \hline
 $B/\bar m$ & $P_{val}$ & $P^{noA}_{val}$& $P^{A}_{val}$
            & $P_{val}$ & $P^{noA}_{val}$& $P^{A}_{val}$
            & $P_{val}$ & $P^{noA}_{val}$& $P^{A}_{val}$ \\
 \hline
 0.10& 0.69&0.01&0.68&0.73 &0.02 & 0.71& 0.75 &0.04 &0.71\\
 0.20& 0.62&0.02&0.60& 0.64  & 0.03 & 0.61   & 0.66      & 0.05 &0.61\\
 0.30& 0.57    & 0.03   & 0.54   &0.58   & 0.04&  0.54& 0.60   & 0.06&0.54\\
 0.40& 0.53    & 0.04   & 0.49   &0.54  &  0.05   & 0.49   & 0.55     &0.07 &0.48\\
 0.50&  0.50   & 0.05   & 0.45   &0.50  &0.05 &0.45& 0.52  &0.07 &0.45\\
 \hline
 \end{tabular}
 \end{center}
 \end{table}
 \begin{figure}
 \includegraphics[width=9cm]{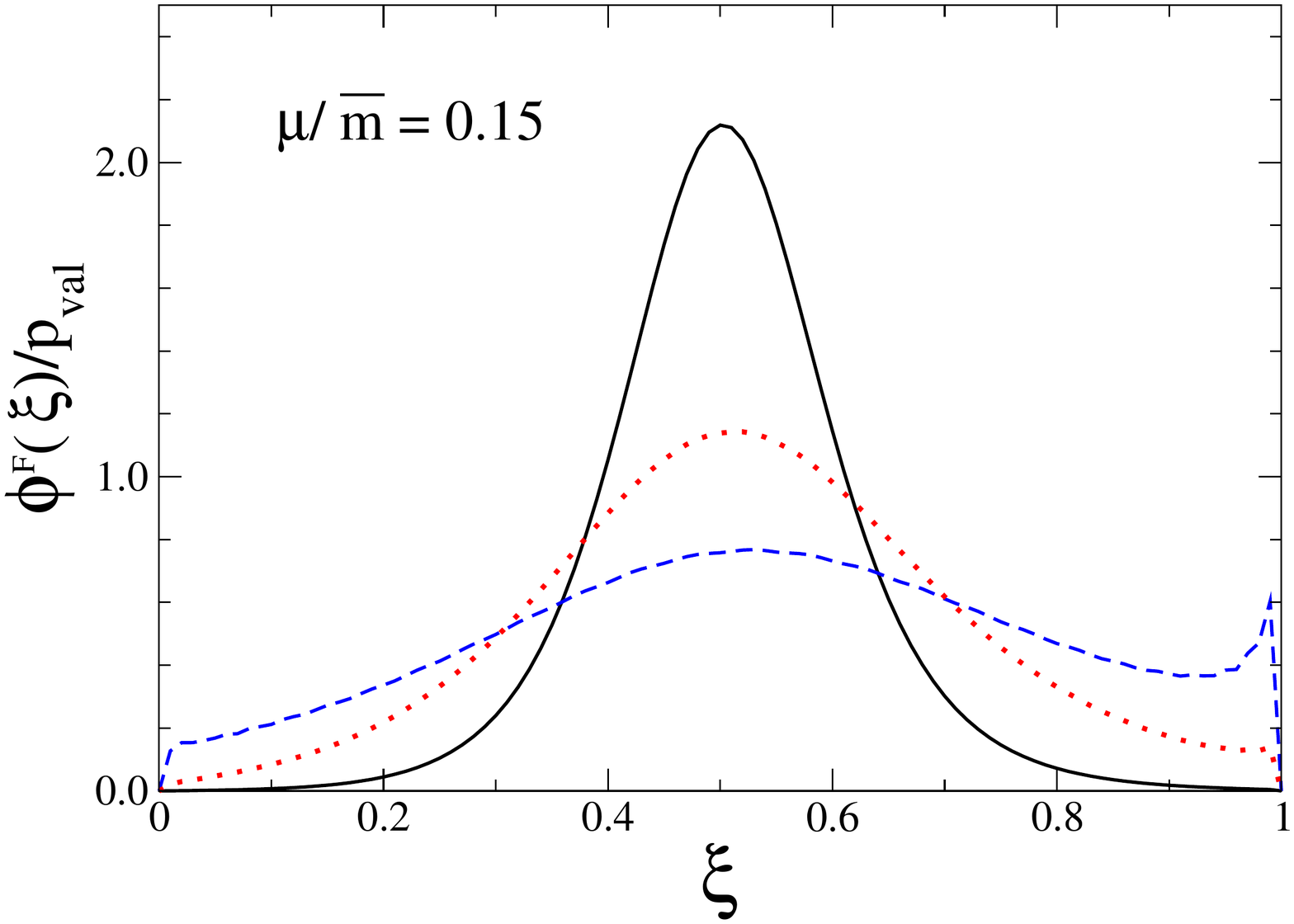} \hspace{-1.8 cm}
 \includegraphics[width=9cm]{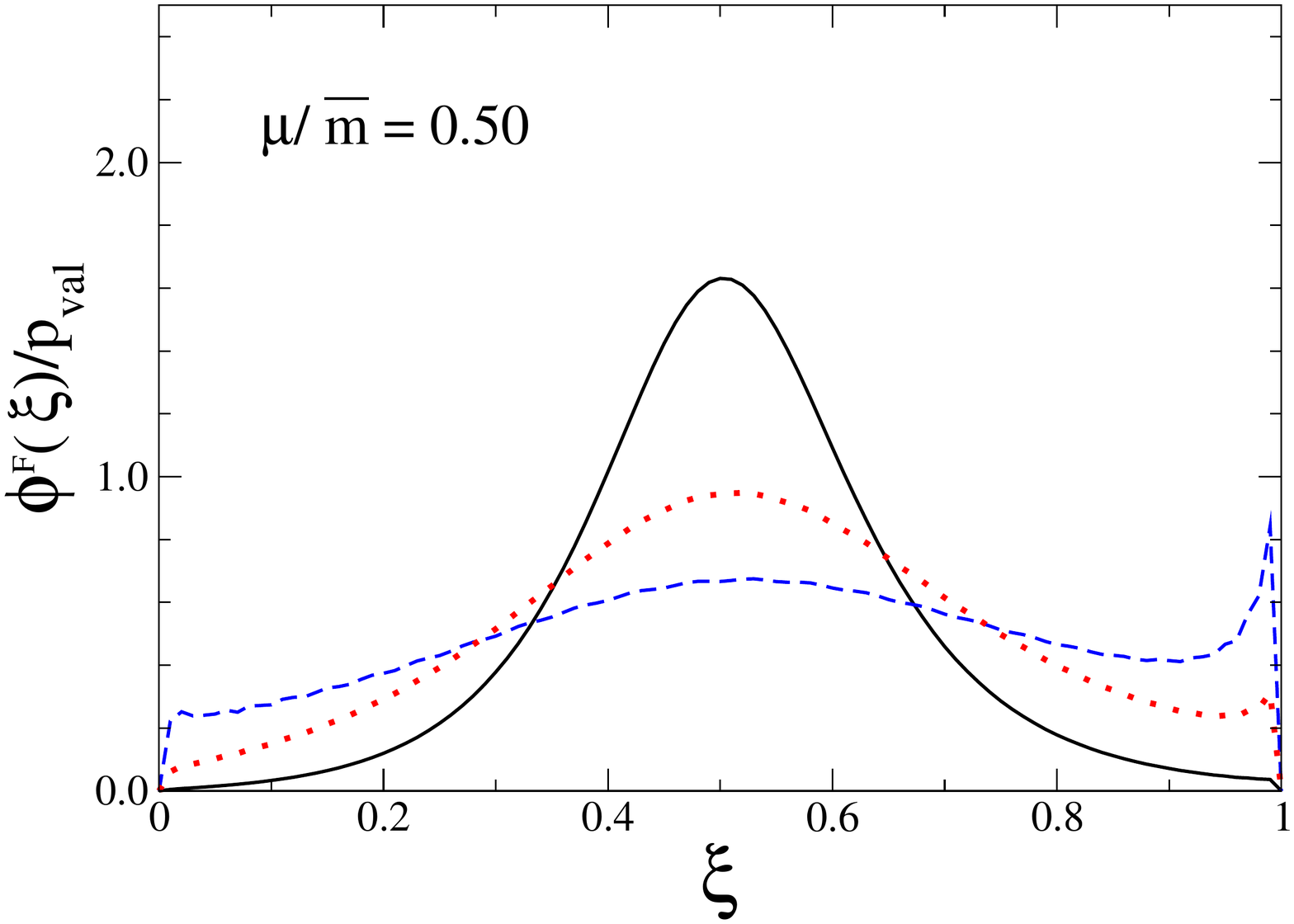}
 \caption{(Color online). Longitudinal LF distributions for  a fermion in the valence
 component for $\mu/\bar m=0.15$ (left panel) and for $\mu/\bar m=0.50$ (right
 panel), in the case of a vector exchange. Solid line: $B/\bar m=0.1$. 
 Dotted red line: $B/\bar m =0.3$. Dashed blue line: $B/\bar m=0.5$.
}
 \label{fig7}
 \end{figure}
 \begin{figure}
 \includegraphics[width=8.7cm]{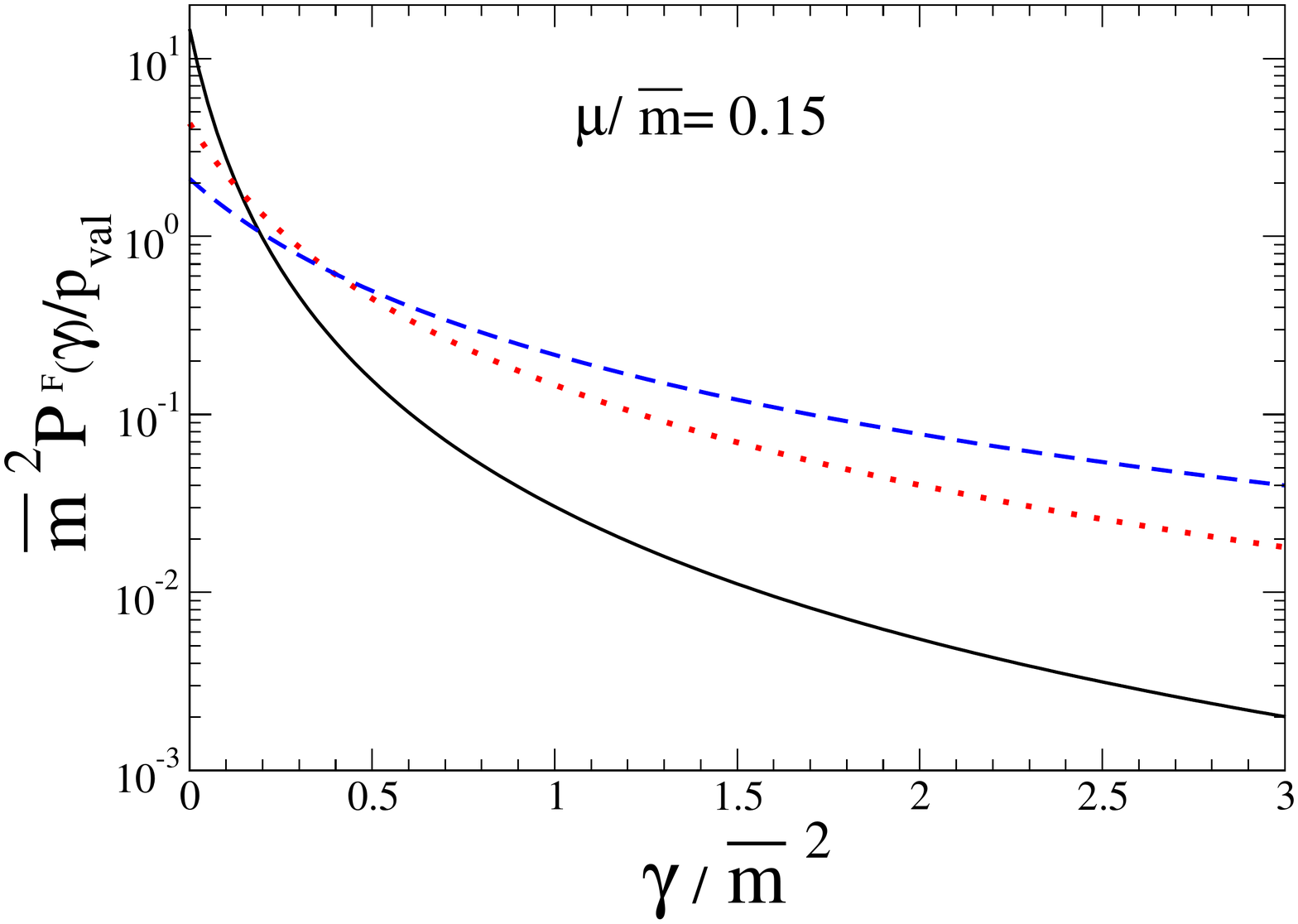} \hspace{-1.3 cm}
 \includegraphics[width=8.7cm]{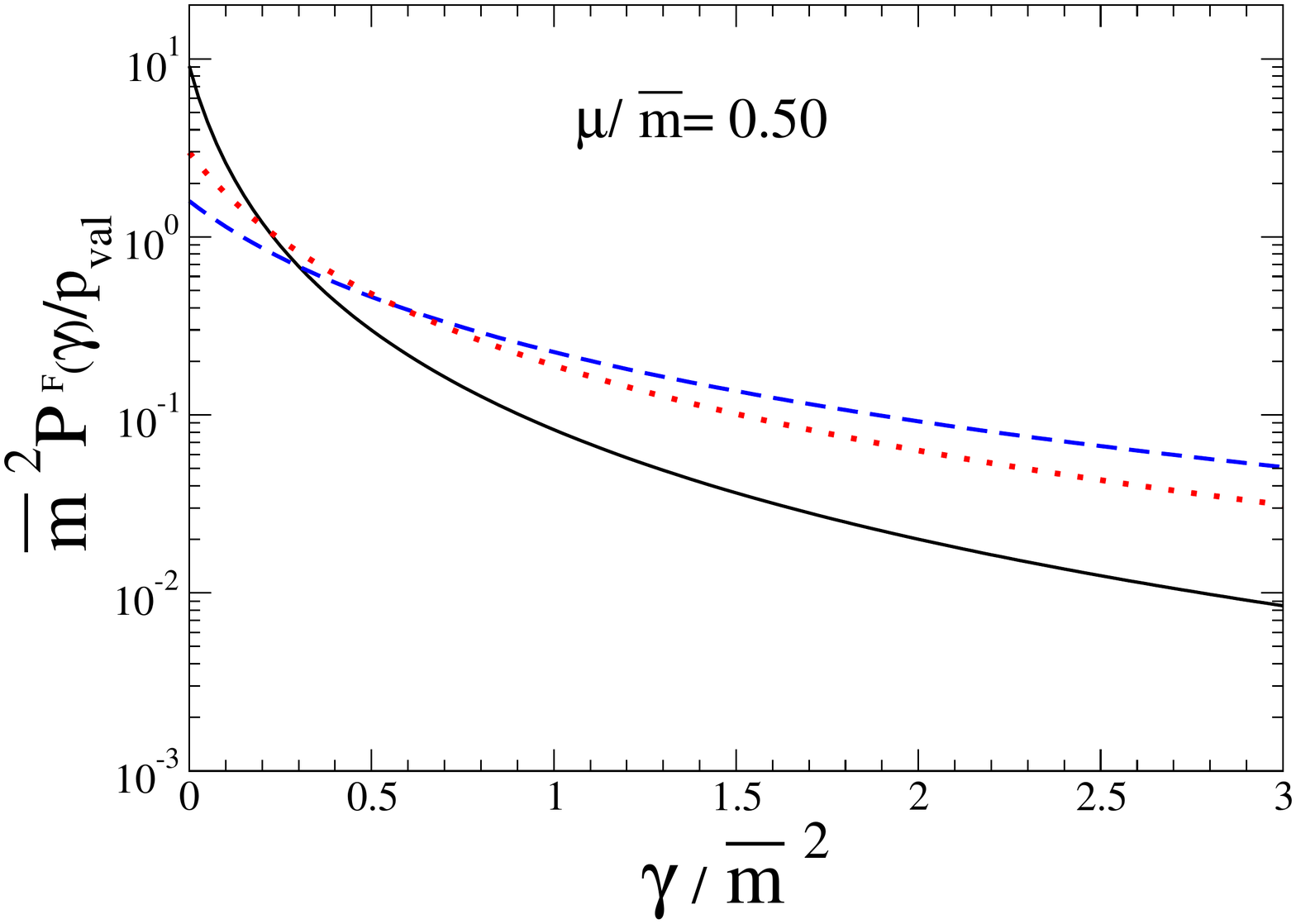}
 \caption{(Color online). Transverse LF distributions for  a fermion in the valence
 component for $\mu/\bar m=0.15$ (left panel) and for $\mu/\bar m=0.50$ (right
 panel), in the case of a vector exchange. Solid line: $B/\bar m=0.1$. 
 Dotted red line: $B/\bar m =0.3$. Dashed blue line: $B/\bar m=0.5$.
 }
 \label{fig8}
 \end{figure}

Figures \ref{fig7} and \ref{fig8} illustrate the 
peculiar pattern of the longitudinal and transverse LF distributions, 
 respectively.  As already noticed for the scalar 
 case, for increasing $B/\bar m$ the size of the system 
 decreases
 and the average kinetic energy of the constituents becomes bigger and bigger.
 This favors the orbital $L=1$ component of the valence wave function, and
 produces an increasing of the non-aligned configuration, mainly close at
 $\xi=0$. As shown in 
 Fig. \ref{fig7}, the longitudinal distribution acquires a particular 
 shape for
 increasing $B/\bar m$, { emphasizing what we have started to see in the scalar-exchange case.
 The calculations show that the dominant contribution for $\xi
 \to 1$ is given by the aligned
 configuration and for $\xi\to 0$ by the anti-aligned one. After recalling that
 for $\xi \to 1$ and $\xi \to 0$ the fermion has a maximal Cartesian component 
 along its spin, one can say that the fermion in both cases has a {\em positive
 helicity. }
 Such a result can be interpreted in the light of the conservation of the angular 
 momentum within the  LF quantum-field theory, or
  equivalently of the helicity conservation for {\em the vector interaction} (see Ref. \cite{Chiu} for a recent
 work elucidating this issue).}
 In conclusion, it is gratifying that the outcome of a non-trivial dynamical  
 calculation
 is in full agreement with the physical expectation from a conservation law.
 
 The transverse distributions show the familiar fall-off that becomes 
 less and less pronounced for
 increasing $B/ \bar m$,  following what we have already learned about the
 increasing of the kinetic energy.
 
 For the sake of completeness, we quote also  the average values of $<\xi>$ and  $<\gamma/\bar
 m^2>$ . In particular, 
  { $<\xi>$  follows a slightly increasing  pattern  from 
 $0.51$ for  $B/\bar m=0.1$ to $0.55$  at
 $B/\bar m=1.0$ (almost irrespective of for $\mu/\bar m$), and  
 $<\gamma/\bar m^2>$   ranges from  $ 0.17$ 
 to  $5.9$ (for $\mu/\bar m=0.50$, the corresponding values are  $0.44$ and 
  $6.8$).
 
  It is worth  noticing    that
 the onset of the helicity conservation should be investigated in more detail, in particular by
 exploring the impact of a non pointlike interaction vertex, i.e. different from the one assumed in
 the present work.
 
 { In Fig. \ref{fig9}, the LF distributions for a fermion-scalar system with different masses of the constituents is
 presented. In order to  start a first survey of a mock nucleon we have chosen a mass ratio 
$m_S/m_F=2$ and a binding energy $B/\bar m =~0.1$ (e.g.,  $M_N/\bar m=1.9$). Also in this case we adopted
 $\mu/\bar m =0.15$ and $\mu/\bar m =0.50$. The corresponding
 values for the coupling constants are $\alpha^V=0.648$ and $\alpha^V=0.898$, 
 respectively, while for 
  the valence probabilities  we have found  $P_{val}=0.75$ and  $P_{val}=0.77$,
  that means a quite large valence component. It is very rewarding
   and of   phenomenological interest to recognize  the signature of the scale invariance in the
    behavior of the  tail of the transverse-momentum distributions. 
    As a matter of fact, extending the calculations for $\gamma/\bar m^2 >40$
    the fall-off  can be described by  $C_1/\gamma^{2.26}$
for  $\mu/\bar m =0.15$, and  $C_2/\gamma^{2.43}$
for  $\mu/\bar m =0.50$, in agreement with the values predicted by the scale invariance analysis 
in Ref. \cite{scaleinv}. Indeed, even if the model is elementary, since both
self-energy and vertex corrections are absent and only the one-vector exchange 
is
taken into account (notice that such an interaction should
govern the tail of the momentum distributions, also in more refined approaches)
this kind of analysis could suggest  the proper 
framework, where a study of the scale
invariance could be started, analyzing the departures from the  evaluations
we are providing.}
 
 \begin{figure}
 \includegraphics[width=8.7cm]{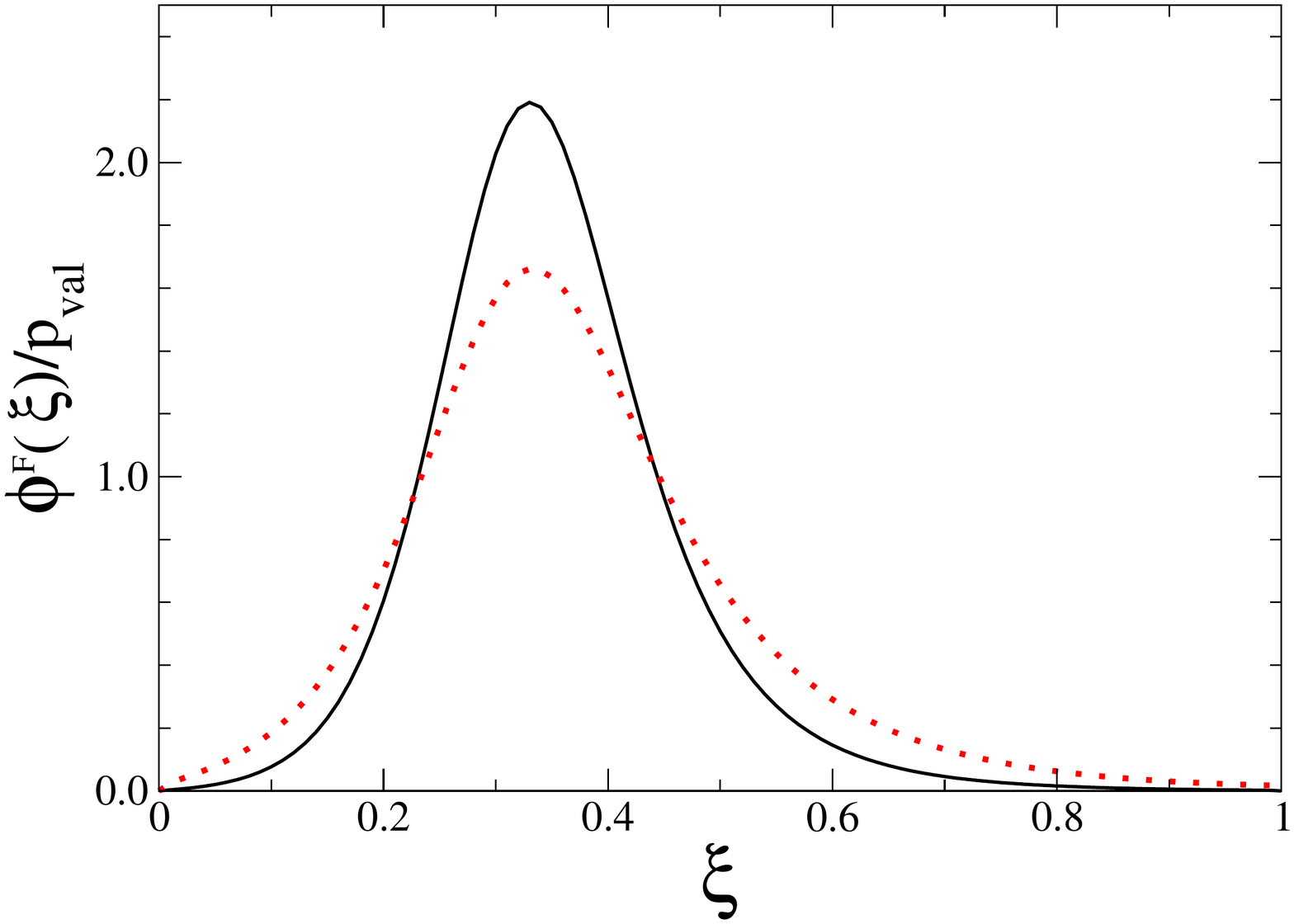} \hspace{-1.3 cm}
 \includegraphics[width=8.7cm]{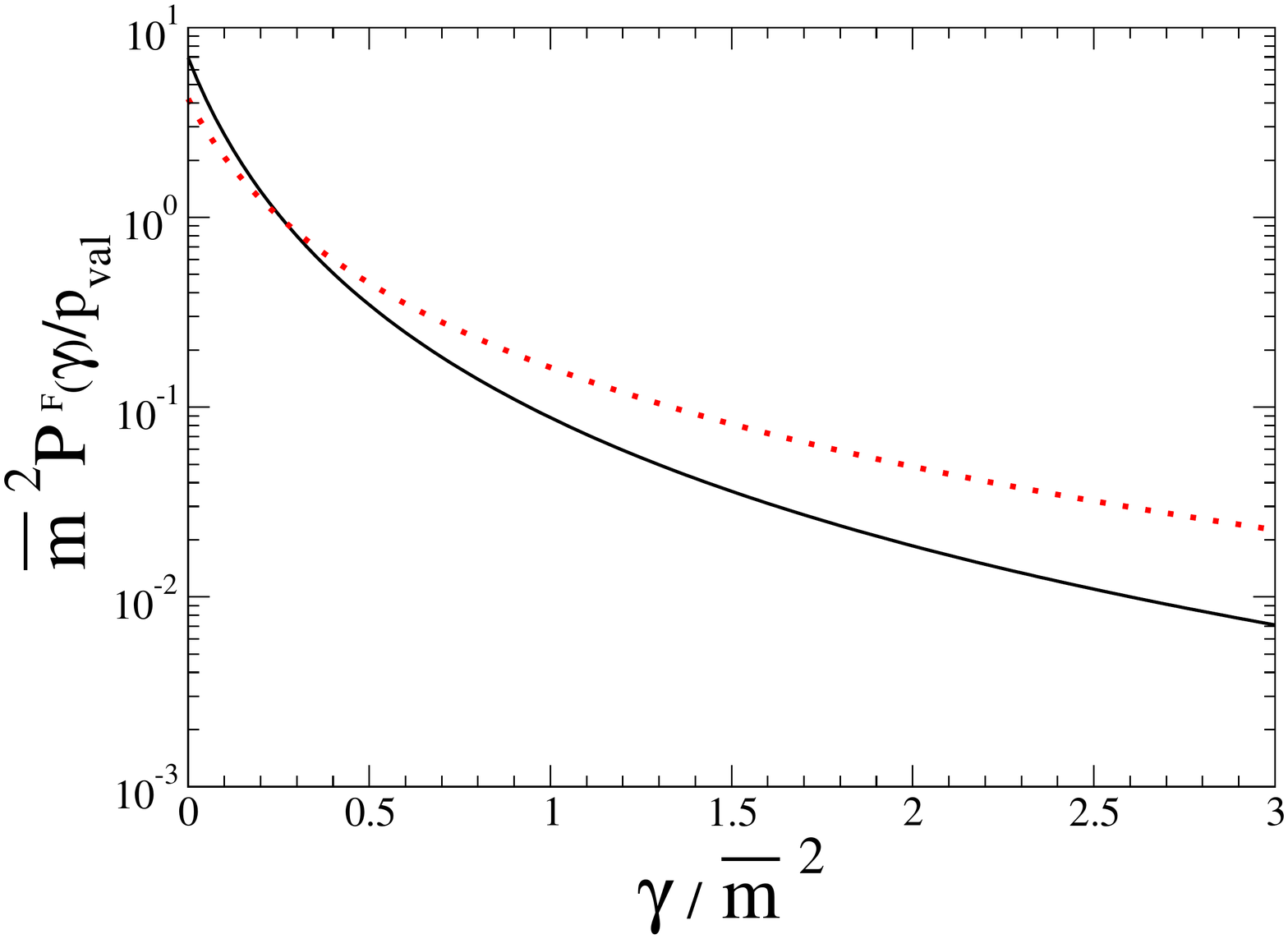}
 \caption{(Color online). Light-front distributions  for  a fermion in the valence
 component of the state $(1/2)^+$, with  a  mass ratio $m_S/m_F=2$ and a binding $B/\bar m =~0.1$. 
 Left panel: longitudinal distribution. Right panel transverse-momentum distribution. Solid line:
  $\mu/\bar m=0.15$. Dotted line:   $\mu/\bar m =0.50$ .
 }
 \label{fig9}
\end{figure}}
 
 \section{Conclusions}
 \label{sect_concl}
 The present study of the ladder  Bethe-Salpeter equation describing an interacting system,
 composed by a fermion and a scalar, has been performed by adopting an approach based on the
 Nakanishi integral representation of the BS amplitude and the light-front framework. In
 particular, two interactions have been applied: i) the scalar interaction and ii) the vector
 one (in Feynman gauge).  
This  model, though simple,  already allows one to start a dynamical investigation 
 of intrinsic features, like  i) in the scalar exchange case, the repulsive 
 effects that
 could pose a  limitation to the use of the  ladder kernel in the
 strong-coupling regime, and ii) in the vector case,
 the effects of the
 helicity conservation, as well as of the
 scale-invariant regime and beyond, that will be investigated elsewhere 
  \cite{scaleinv}. 

 Moreover, one has to emphasize 
 the absence of the
 exchange symmetry, working for the two-scalar and two-fermion systems.
 
 The coupling constants for assigned masses of the interacting system have
  been obtained as
 an outcome of an eigenvalue problem, formally deduced from the initial BSE, and compared with
 the corresponding results where the BSE is solved after introducing a 
 Wick rotation. The
 agreement, as in the case of two-scalar \cite{CK2006,FSV2} and 
 two-fermion systems
 \cite{CK2010,dFSV1,dFSV2}, is very good, and adds more and more confidence 
 in the adopted
 approach. 
 
 To conclude, a benefit of any technique able to solve the BSE in Minkowski space is the direct access to the LF
 distributions,  that have been shown in Figs. \ref{fig4}, \ref{fig5}, \ref{fig7}, \ref{fig8}
  and \ref{fig9}.  {For the equal-mass case,} the very peculiar behavior
 of the longitudinal distributions, with a particular emphasis for the vector interaction, has
 allowed us to point out the nice role played by the two possible configurations of the
 constituent and system spins. In the case of the vector interaction, 
    when states of large momentum are more and more populated for   
 increasing values of the coupling constant, 
 the fermion longitudinal distribution sizably cumulates  close to $\xi \to 0$ and  $\xi\to 1$,
   yielding
  a clear evidence of the action of the helicity conservation law.  
  Furthermore,
   it should be interesting 
  to notice that the {\em relative 
 weight} of
 the orbital $L=1$ (anti-aligned configuration) with respect to the $L=0$ one (aligned configuration) 
 is around 10\% for both 
 interactions, at $B/\bar m=0.5$, with  
 an increasing behavior for   increasing binding energies. 
 { The unequal-mass  case with a vector-exchange interaction,  i.e. 
  a
 mock nucleon composed by a quark and a scalar {point-like} diquark, with mass ratio 
 $m_S/m_F=2$, yields the possibility to investigate the extent to which
  the scale
 invariance could affect the hadron dynamics. Indeed,
  our calculations, that necessarily lead to a scale-invariant behavior of the
  transverse-momentum distribution (cf Fig. \ref{fig9}), should be considered 
   as a reference line 
 for more refined  phenomenological 
 studies. Finally, we should point out  that    
   massless fermion-scalar systems (e.g.,  a 
 ghost-quark  bound state as in Ref. \cite{Alkofer:2011pe})   can be addressed
  only  by introducing a new scale other than the masses of constituents and
  quanta, like the one  associated to  
 an extended interaction vertex (cf the results in Refs. \cite{dFSV1,dFSV2}). 
 This very interesting study has to be
postponed to another work, where, both vertex and self-energy corrections 
have to be carefully treated.}

\begin{acknowledgments} 
 TF thanks Funda\c c\~ao de Amparo \`a Pesquisa do Estado 
 de S\~ao Paulo (FAPESP)  Thematic  grants  no.   13/26258-4  
 and  no.   17/05660-0.   JHAN thanks FAPESP grant  no.  2014/19094-8 and no. 
 2017/14695-1; 
 TF thanks Conselho Nacional  de  Desenvolvimento  Cient\'ifico  e  
 Tecnol\'ogico  (Brazil)  and  Project  INCT-FNA  Proc. No.  464898/2014-5. 
  This study was financed in part by Coordena\c c\~ao de Aperfei\c coamento 
  de Pessoal de N\'ivel Superior (CAPES) - Finance Code 001.
\end{acknowledgments} 
\appendix

\section {Coefficients for the scalar-boson exchange}
\label{scal}
The coefficients in Eq. \eqref{bse_coup} for the exchange of a scalar-boson are given by
\be { \cal {C} }_{ 11 }=\frac{1}{2}M+{ m }_F \nonu
{ \cal {C} }_{ 12 }={\spro{k'}{p}\over M }-{1 \over M^2 k^ 2 -\spro{k}{p}^2 }
 \left\{  {\spro{k'}{p}\over M }~ k^2\left[ M^2b +  
 \spro{k}{p} \right]
 \nonub -  (k'\cdot k) ~M\left[\spro{k}{p}
b+ k^2 \right] \right\}
\nonu
{ \cal {C}}_{ 21 }=M \nonu
{ \cal {C} }_{ 22 }={M \over M^2 k^ 2 -\spro{k}{p}^2 }
 \left\{\spro{k'}{p}\left[ k ^2 +
b \spro{k}{p}\right]
 -  (k'\cdot k) \left[M^2b+\spro{k}{p}\right]\right\}\nonu
\label{coeff_scal}
\ee
where $$b=\frac{1}{2}-{ m _F\over M }$$

The non-vanishing coefficients in Eq. \eqref{Bcoeff} are given by
\be
  c^{(0)}_{ 11 }={M\over 2}a 
  \nonu
  c^{(0)}_{ 12 }=-\frac { z'v }{ 2 }~{M\over 2}a 
 -\left( 1-v \right) ~\frac { 1 }{ M }
  \left( \gamma +\frac { zM^2 }{ 4 }  \right) ~, 
~~ ~~
 c^{(1)}_{ 12 }= {\left( 1-v \right)\over 2}   \left( 1-z \right)
  \nonu 
  c^{(0)}_{ 21 }=M 
  \nonu
  c^{(0)}_{ 22 }=-M\frac { z'v }{ 2 } -\left( 1-v \right) ~Mb 
\label{Ccoeffs}\ee
with $$a= 1+2{m_F\over M}$$
\section {Coefficients for the vector-boson exchange}
\label{vect}
The coefficients in Eq. \eqref{bse_coup} for the exchange of a vector-boson are given by
\be
 {\cal C}_{11}
={M^2 \over 2} a -
k^2
-\spro{k}{p}-\spro{k'}{p}+{M^2 \over 2\left[k^2 M^2 -
\spro{k}{p}^2\right]}
\nonu \times \left\{(2-a)\left[(k'\cdot p)k^2 -\spro{k}{k'} \spro{k}{p}\right]
+ 2{k^2\over M^2}\left[\spro{k'}{p}\spro{k}{p} 
- \spro{k}{k'}M^2\right]\right\}
\nonu \nonu
{\cal C}_{12}=
~-{a\over 2}k'^2 -\spro{k'}{k}+(a-1)\spro{k'}{p}+{M^2 \over 2\left[k^2 M^2 -
\spro{k}{p}^2\right]}
\nonu \times ~
\left\{ \left[(2-a)- 2{k^2\over M^2}\right]~
\left[ \spro{k'}{p} k^2-\spro{k'}{k}\spro{k}{p} \right] 
 + a {k^2 \over M^2}\left[\spro{k'}{p}\spro{k}{p} - \spro{k'}{k} M^2 
 \right]\right\}
\nonu
\nonu
 {\cal C}_{21}
=  (2-a ){ M^2\over 2}
+M^2- {M^2 \over 2\left[k^2 M^2 -
\spro{k}{p}^2\right]}
\nonu \times \left\{(2-a)\left[\spro{k'}{p}\spro{k}{p}-\spro{k}{k'}M^2 \right]
+2\left[ \spro{k'}{p} k^2 -\spro{k}{k'}\spro{k}{p}
  \right]\right\}
\nonu \nonu
 {\cal C}_{22}
=
-
 k'^2- \spro{k'}{k}+2\spro{k'}{p} - {M^2 \over 2\left[k^2 M^2 -
\spro{k}{p}^2\right]}
\nonu \times~\left\{
(2-a)\left[\spro{k'}{p}\spro{k}{p}-\spro{k}{k'}M^2 \right]
+ \left[a - 2{\spro{k}{p}\over M^2}\right]~\left[ \spro{k'}{p} k^2-\spro{k}{k'}\spro{k}{p}
   \right]\right\}
\nonu\ee
where
\be
a=1 +{2 m_F\over M}
\ee
The coefficients in Eq. \eqref{Bcoeff} are given by
\be
c^{(0)}_{11}= {M^2\over 2} \left[a 
+v z'{a\over 2} +(2-v) {z\over 2} \right]
+(2-v)\gamma \nonu
c^{(1)}_{11}= -{M \over 2} (2-v)\left(1 -z \right)
\nonu
c^{(0)}_{12}={ a\over 2}\left\{
\gamma(1-v) (2+v)+2v(\gamma '+ \kappa^2)+2(1-v) \mu^2  \phantom{1\over 2}
\nonub+{M^2\over 4}
(v  z' -2)[z
-v(z-z')] \right\}
+\left(\gamma
+z{M^2\over 4} \right)~\left(1-v-z'{v\over 2}\right)
\nonu
c^{(1)}_{12}={ M\over 2}~\left\{a(1-v)\left[  1+z +(z-z'){v\over 2}
\right]
-{(1-z)} ~\left(1 -v-z'{v\over 2}\right)\right\}
\nonu
c^{(0)}_{21}={M^2\over 2} \left[(2-v)(2-a) + 2 + z'v\right]
\nonu
c^{(0)}_{22}=\gamma(1-v)(2+v) +2v(\gamma '+ \kappa  ^2)+2(1-v){ \mu  }^2
\nonu +{M^2\over 2 } \left[(1-v) \left( zz'{v\over 2} -z +2-a\right)
+z'{v\over 2} ~( z' v -4 +a)\right]
\nonu
c^{(1)}_{22}
={M \over 2}~(1-v)
\left[v(z-z')+2(1+z)\right]
\ee
\section {The  normalization of the BS amplitude}
\label{normbsa}
The normalization of the  BS amplitude is obtained by applying the standard
expression  illustrated, e.g.,  in Ref. \cite{Lurie}, but adapted to the
fermion-scalar case. In particular, recalling that we disregard the
self-energy effects, one has the following 
 normalization constraint
\be
\int {d^4 q\over (2\pi)^4} \int {d^4 k\over (2\pi)^4} ~\bar \Phi^\alpha_p(q,J'_z)
~ \left.{\partial \over \partial p_\mu} \Bigl[G^{-1}_0(k,p) (2\pi)^4 \delta^4(q-k)-i 
{\cal K}(q,k,p)\Bigr]\right|_{p^2_{on}}
\nonu
\times ~
\Phi^\beta_p(k,J_z)=i2 p^\mu~\delta_{J'_z,J_z}~\delta_{\alpha\beta}
\label{norm}\ee
where
\be
G^{-1}_0(k,p)=-\left({p^2\over 4}+k^2-p\cdot k -m^2_S\right)~ \left(\psla p/2
+\psla k -m_F\right)
\ee
Depending upon the actual expression of the interaction kernel in ladder
approximation, one has or not a contribution for the derivative of ${\cal K}$.
Finally, it is worth mentioning that in ladder approximation the normalization 
amounts to the charge normalization.
\subsection{Scalar-exchange kernel}
Let us consider the scalar-exchange case. 
In ladder approximation the interaction kernel ${\cal K}$ does not depend upon $p$, and
therefore does not contribute to the derivative in Eq. \eqref{norm}.
By inserting the BS amplitude as given in Eq. \eqref{bsa}, one remains in the
scalar case with
\be
 \int {d^4 k\over (2\pi)^4} ~\bar U^\alpha(p,J'_z)
\left[ \phi_1(k)+ {\psla k\over M} \phi_2(k)\right]\nonu
\times
~\left[\left(m^2_S -{3 \over 4} M^2 -k^2 +2p\cdot k   \right)
\psla p +\left(2 p \cdot k -M^2\right)~(\psla k-m_F)\right]\nonu
\times~
\left[ \phi_1(k)+ {\psla k\over M} \phi_2(k)\right] U^\beta(p,J_z)=i 4 M^2~
\delta_{J'_z,J_z}~\delta_{\alpha\beta}
\ee
In the CM frame, after multiplying with the proper spinors and summing over
$J'_z$ and $J_z$, one gets 
\be
 \int {d^4 k\over (2\pi)^4} ~{(\psla p+M)\over 2M}
\left[ \phi_1(k)+ {\psla k\over M} \phi_2(k)\right]\nonu
\times
~\left[\left(m^2_S -{3 \over 4} M^2 -k^2 +2p\cdot k   \right)
\psla p +\left(2 p \cdot k -M^2\right)~(\psla k-m_F)\right]\nonu
\times~
\left[ \phi_1(k)+ {\psla k\over M} \phi_2(k)\right] {(\psla p+M)\over 2M}=i 4 M^2~
~{(\psla p+M)\over 2M}
\ee
where $$\sum_{J_z}  U(p,J_z)\bar U(p,J_z)={(\psla p+M)\over 2M}$$
and $\bar U(p,J_z) U(p,J_z)=1$.

Finally, one  evaluates the traces, takes care of the NIR for $\phi_i$ (Eq.
\eqref{phi}) and performs the 4D integration by exploiting
standard tricks (see, e.g., \cite{FSV2}), obtaining the following normalization constraint
\be
{ M\over (8\pi)^2}
\int_{\gamma_{min}}^\infty d\gamma'' \int_{\gamma_{min}}^\infty d\gamma' \int_{-1}^1 dz''\int_{-1}^1 dz'
\int_0^1 dv~ v^2 (1-v)^2~{1\over A^4}
~\Bigl\{ N_{11}~g_1(\gamma'',z'';\kappa^2)~g_1(\gamma',z';\kappa^2)
\nonu +N_{12}~g_1(\gamma'',z'';\kappa^2)~g_2(\gamma',z';\kappa^2)
+N_{22}~g_2(\gamma'',z'';\kappa^2)~g_2(\gamma',z';\kappa^2)
\Bigr\}
=1
\label{norms}\ee
where
\be
N_{11}={\cal C}+{A\over 2M^2}
\nonu
N_{12}=-2\left[\lambda~
{\cal C}
+{A\over 2M^2}  \left(1+ 3 \lambda+
{m_F \over M}\right)
\right]
\nonu
N_{22}=\lambda^2~{\cal C}+{3\over 2} {A^2\over M^4} -{A\over 2M^2}
\left[2{m_F\over M}
 -{m^2_S\over M^2} 
+{3 \over 4}  +{3}\lambda~\left(2{m_F\over M}  +  1\right)+
3 \lambda^2  \right]
\ee
with
\be
{\cal C}={3\over 2}~\left[ 
{m^2_S\over M^2} -{3 \over 4}  +{m_F\over M}
 -\lambda \left ( 1- 2{m_F\over M}\right) + \lambda^2\right]
\nonu
A=\kappa^2(1-4\lambda^2)   
+v\gamma'' +(1-v)\gamma'
+(2 \lambda \bar m -\Delta)^2 ~~,
\nonu
\lambda= [v z''+(1-v)z']/2~~.
\ee
\subsection{Vector exchange kernel}
In the vector-exchange case, the interaction kernel ${\cal K}$ acquires a 
dependence upon
the total momentum $p$, and therefore one has 
\be
{\partial\over \partial p_\mu} \left[i{\cal K}(k,k',p)\right]=
-i \lambda^v_S \lambda^v_F{\gamma^\mu\over (k-k')^2-\mu^2+i\epsilon}~~~.
\label{deriK}\ee
After performing  steps similar to the ones done for the scalar exchange, 
one obtains a contribution
 generated by the  derivative in Eq. \eqref{deriK}, that has to be added to the one shown in the
lhs of Eq. \eqref{norms}. The actual form of this new contribution is
\be
{1\over 2  M^2 ~ (4\pi)^3} {\lambda^v_S \lambda^v_F\over 8\pi}
\int_{\gamma_{min}}^\infty d\gamma'\int_{-1}^1 dz'~
\int_{\gamma_{min}}^\infty d\gamma\int_{-1}^1 dz~  ~
\int_0^1 dv ~v^2~\int_0^1 d\xi
{\xi^2~(1-\xi)\over
\left\{D_3-
i\epsilon\right\}^3}
\nonu \times ~\left\{Ma g_1(\gamma',z';\kappa^2) g_1(\gamma,z;\kappa^2)
-Mbg_1(\gamma',z';\kappa^2) g_2(\gamma,z;\kappa^2)
\right. \nonu \left. +g_2(\gamma',z';\kappa^2) g_2(\gamma,z;\kappa^2)
\left[z'vb {M\over 4} +  {1 \over 2 aM}
(1-v)\left( D_3 +b^2{M^2\over 2} \right) \right]\right\}\ee
where
\be
D_3={M^2\over 4} {b^2}+a~\xi(\gamma+\kappa^2)
+a~(1-\xi)\left[v^2z^{\prime 2}{M^2\over 4}
+v(\gamma'+\kappa^2)+(1-v)\mu^2\right]
\nonu
a=\xi +(1-\xi)v(1-v)~~,  \quad b= \xi z+(1-\xi) v(1-v)z'
\ee

\section{The valence component}
\label{app_vale}
In this Appendix, the relation between the BS amplitude and the valence
component of the fermion-scalar interacting state is discussed with some detail.

For illustrative purpose, we assume a scalar exchange and write  
the Fock expansion of the fermion-scalar interacting system  as
follows
\be
|\tilde p;M,JJ_z; \pi;\rangle= 2 (2\pi)^3 \sum_{n\ge 2}\sum_{n_F=1}^{n-1}
\sum_{\{\sigma_i \}_{n_F}}
 ~\int [d\xi_{i}] 
\int [d{\bmm \kappa}_{i\perp }] \nonu \times ~
\psi^{J\pi}_n(\{\xi_{i}\}_n ;\{\kappa_{i \perp }\}_n;\{\sigma_i\}_{n_F} ;J_z)
 ~|\{\xi_i p^+,{\bmm \kappa}_i+\xi_i{\bf p}_\perp\}_n;  \{\sigma_i\}_{n_F}\rangle
\label{fock1}\ee
where the integration symbols mean
\be
\int \big[d \xi_i\big] \equiv \prod_{i=1}^n \int {d\xi_i\over 2 ~(2\pi) \xi_i } 
\,\delta\left(1- \sum_{j=1}^n\xi_j\right)
~~, \nonu
\int [d{\bmm \kappa}_{i\perp }]\equiv 
\prod_{i=1}^n \int {d{\bmm \kappa}_{i\perp }
\over (2\pi)^2}~\delta^2\left( \sum_{j=1}^n{\bmm \kappa}_{j\perp }\right)
\ee
In Eq. \eqref{fock1}, the generic Fock state  contains   $n_F$ and $n_S$  
fermionic and scalar constituents, respectively, and    $n_E$   exchanged 
bosons. It is given by (recall that $n=n_F+n_S+n_E$)
\be
|\{\tilde q_i\}_n;\{ \sigma_i\}_{n_F}\rangle= 
(2\pi)^{3n/2}~{1\over \sqrt{n_F!}}~{1\over \sqrt{n_S!}}
~{1\over \sqrt{n_E!}}~\Pi_{j=1}^{n_S}\sqrt{ 2 q^+_j} ~a^\dagger(\tilde q_{j})
\nonu \times ~
\Pi_{\ell=1}^{n_E} \sqrt{ 2  q^+_{\ell}}~c^\dagger(\tilde q_{\ell})
~\Pi_{r=1}^{n_F} \sqrt{ 2  q^+_{r}}~
b^\dagger(\tilde q_r,\sigma_r)~|0\rangle
 \label{focks}
 \ee
In the above equation,  $a^\dagger(\tilde q_j)$ and
 $c^\dagger(\tilde q_\ell)$ are the creation 
 operators of constituent scalars and exchanged
 bosons, respectively, while the operators $b^\dagger(\tilde q_r,\sigma_r)$ create fermions.
 In Eq. \eqref{focks}, the symbols $\{{\cal O}_i\}_{\ell}$ indicate  
 ${\cal O}_1,~{\cal O}_2,~\dots
 ~\dots~,
 {\cal O}_\ell$. The
 normalization reads
 \be
 \langle  \{\sigma'_i\}_{n_F}; 
 \{\tilde q'_i\}_{n} |\{\tilde q_i\}_{n}  \{\sigma_i\}_{n_F}\rangle
 =(2\pi)^{3n}~n_S!n_E! n_F!\prod_{\ell=1}^{n}~2 q^+_\ell 
 \delta^3\Bigl(\tilde q_\ell- \tilde q'_\ell\Bigr)~\delta_{\sigma_\ell, \sigma'_\ell}
\nonu \ee 
 Notice that if $n_F$ is odd (even) then  $J=(2m+1)/2$ ( $J=2m$).
 
 In Eq. \eqref{fock1}, the functions $\psi^{J\pi}_n$ are the 
 {\em LF wave amplitudes} (aka LF wave functions), and 
 the first one, i.e. the amplitude of the Fock state with the lowest number 
 of
 constituents and no exchanged boson, is the {\em valence wave function}.

The normalization of the full interacting state is taken to be
\be
\bigl\langle \pi; J'_z,J,M,\tilde p' \big\vert \tilde p, M,J,J_z; \pi\bigr\rangle 
= 2 p^+~(2 \pi)^3\delta^3
\bigl(\tilde p' - \tilde p\bigr)~
 ~\delta_{J'_z,J_z}~
\bigl\langle \pi;J_z,J,M\big\vert   M,J,J_z;\pi\bigr\rangle 
\nonu \label{fock2}\ee
where $\bigl\langle \pi;J_z,J,M\big\vert   M,J,J_z;\pi\bigr\rangle $ is the
normalization of the intrinsic part of the  state.

On the other hand, from Eq. \eqref{fock1}, one can write
\be
\bigl\langle \pi; J_z,J,M,\tilde p' \big\vert \tilde p, M,J,J_z;\pi\bigr\rangle
=  
[2 p^+ (2\pi)^3]^2 \sum_{n \ge 2} \sum_{\{\sigma_i\}_{n_F}}~
  \prod_{i=2}^n\int  {d^3 \tilde q_i\over 2q^+_i (2\pi)^3}~
 \delta^3\Bigl(\sum_{i=1}^n \tilde q_i- \tilde p\Bigr)
 \nonu \times~
 \delta^3\Bigl(\sum_{i=1}^n \tilde q_i- \tilde p'\Bigr)~
 \big\vert
 \psi^{J\pi}_n(\{\xi_{i}\}_n ;\{{\bf q}_{i \perp }\}_n;\{\sigma_i\}_{n_F} ;J_z)\big\vert^2
 = 
 \nonu
 =2p^+ (2\pi)^3 \delta^3\bigl(\tilde p' - \tilde p\bigr)~ 2~(2\pi)^3
 ~\sum_{n\ge 2}\sum_{\{\sigma_i\}_{n_F}}~\int \big[d \xi_i\big] \left[d^2
{\bf q}_{i\perp }\right]\,
 ~\big\vert 
\psi^{J\pi}_n(\{\xi_{i}\}_n ;\{{\bf q}_{i \perp }\}_n;\{\sigma_i\}_{n_F} ;J_z)
\big\vert^2 
\nonu \label{fock3} \ee
 If the intrinsic state is normalized, then combining Eqs. \eqref{fock2} 
 and \eqref{fock3},  one can deduce the  following normalization 
 of the LF wave functions, $\psi^{J\pi}_n
$,  viz
\be
2 (2\pi)^3\sum_{n\ge 2}  \sum_{\{\sigma_i\}_{n_F}}~\int \big[d \xi_i\big] 
\left[d^2 {\bf q }_{i\perp }\right]
\,\left\vert \psi^{J\pi}_n(\{\xi_{i}\}_n ;\{{\bf q}_{i \perp }\}_n;\{\sigma_i\}_{n_F} ;J_z)
 \right\vert^2 = 1 \nonu
\label{LFWFnorm}
\ee
Such a normalization of the LF amplitudes is the key point for introducing a
probabilistic description for a relativistic interacting state.
In particular,  
the probability to find the valence 
component in the bound state with  $J=1/2$ and third component $J_z$
is given by 
\be
P_{val}= 2~(2\pi)^3~ \sum_{\sigma_1}\int {d\xi_1\over 2 ~(2\pi) \xi_1 } \int {d\xi_2\over 2 ~(2\pi) \xi_2 } ~\delta(1 -\xi_1-\xi_2)~\times
\nonu
\int
{d^2 {\bmm \kappa}_{1\perp }\over (2\pi)^2}\int
{d^2 {\bmm \kappa}_{2\perp }\over (2\pi)^2}~ \delta^{2} 
\Bigl({\bmm \kappa}_{1\perp }+{\bmm \kappa}_{2\perp }\Bigr)~
\left\vert \psi^{J\pi}_{n=2}(\xi_1 ,\xi_2  ;
{\bmm \kappa}_{1\perp},{\bmm \kappa}_{2\perp} ;\sigma_1 ;J_z)\right\vert^2=\nonu
=  {1 \over (2 \pi)^3}~\sum_{\sigma_1}\int {d\xi \over 2 ~ \xi (1-\xi)} 
\int
d^2 {\bmm \kappa}_{\perp }~
\left\vert \psi^{J\pi}_{n=2}(\xi  ;{\bmm \kappa}_\perp;\sigma_1 ;J_z)\right|^2
\label{fock4}\ee
where the notation has been simplified, putting $\xi=\xi_1$ and 
$\bmm{\kappa}_{\perp }=\bmm{\kappa}_{1\perp }$.

Notice that the valence probability is equal for $J_z=\pm 1/2$.

To establish the relation between 
 $\psi^{J\pi}_{n=2}(\xi  ;{\bmm \kappa}_\perp;\sigma_1 ;J_z)$ and the 
 BS amplitude
  (cf, e.g. \cite{FSV2}) one has to project 
 the Fock expansion in Eq. \eqref{fock1}
 as follows
\be
\langle \tilde q_2 \tilde q_1 \sigma_1  \big\vert \tilde p,  M,J,J_z;\pi\bigr\rangle=
\nonu
= (2\pi)^3 2\sqrt{q^+_1 q^+_2}~2 (2\pi)^3 \sum_\sigma 
\int {d\xi\over (2 \pi)^2 4\xi (1-\xi)} 
~ \int {d{\bmm \kappa}_{\perp}\over (2\pi)^4}
~ \psi^{J\pi}_{n=2}(\xi  ;{\bmm \kappa}_{\perp};\sigma ;J_z)
\nonu
\times ~
\langle 0| a(\tilde q_2) ~b(\tilde q_1,\sigma_1)~
b^\dagger(\tilde k,\sigma)~a^\dagger(\tilde k')|0\rangle~(2\pi)^3~2 p^+\sqrt{\xi (1-\xi)}
=
\nonu
=  2(2\pi)^3~{p^+} ~\delta(q^+_1+ q^+_2 -p^+) ~\delta^2({\bf q}_{1\perp}
+{\bf q}_{2\perp}-{\bf p}_\perp)~ \psi^{J\pi}_{n=2}(\xi ;{\bf q}_{1\perp};\sigma ;J_z)
\label{mt1}\ee
where $\tilde k\equiv \{ \xi p^+, {\bmm \kappa}_\perp+\xi {\bf p}_\perp\}$ and 
$\tilde k'\equiv
\{(1- \xi) p^+,- {\bmm \kappa}_\perp+(1-\xi){\bf p}_\perp\}$.
Following Yan \cite{Yan}, the creation and annihilation operators have to be 
defined in 
terms of the independent degrees of freedom. In particular,
 the fermionic operators are expressed through the {\em good component}
of the field, $\psi^{(+)}(\tilde x,x^+)$, on the hyperplane $x^+=0$,
i.e. $\Lambda^+
\psi(\tilde x,x^+=0)$ with $\Lambda^+=\gamma^0\gamma^+/2$. Hence, one gets 
\be
\psi^{(+)}(\tilde x,x^+=0)
=\sum_\sigma \int  d k^+\sqrt{ m\over  k^+}
\int\frac{d^2 \bmm{k}_\perp}{ (2 \pi)^{3/2}}
~\theta(k^+ )
\nonu \times ~\left[b(\tilde
k,\sigma)e^{-i\tilde k\cdot \tilde x}~u^{(+)} (\tilde k,\sigma)+
d^\dagger(\tilde k,\sigma)e^{i\tilde k\cdot \tilde x}~v^{(+)} (\tilde
k,\sigma)\right]
\nonu\ee
where the LF spinors (recall that $\bar
u u =1$, since in  Appendix \ref{normbsa} the BS norm has been evaluated by using 
$\sum_\sigma ~ \bar u_\sigma u_\sigma=(\psla p +M)/2M$
) are such that 
\be
u^{(+)} (\tilde k,\sigma)=\Lambda^+~u(\tilde k,\sigma)
\nonu
u^{(+)\dagger} (\tilde k,\sigma)~u^{(+)} (\tilde k,\sigma)= 
{1\over 2}\bar u(\tilde k,\sigma)~\gamma^+u(\tilde k,\sigma)= {k^+\over 2m}~~~.
\ee
For instance, the annihilation operator is 
\be
(2\pi)^{3/2} ~ \sqrt{q^+\over m} ~b(\tilde q,\sigma')=\int d\tilde x ~e^{i \tilde q \cdot \tilde x}
  ~ u^{(+)\dagger}(\tilde q,\sigma') 
  \psi^{(+)}(\tilde x,x^+=0)\nonu
\ee
For the scalar case, where there is not the issue of the independent degrees of
freedom (see also \cite{FSV2}), the field is 
\be\varphi(\tilde x,0)=  \int  \frac{d k^+}{\sqrt{ 2 k^+}}
\frac{d^2 \bmm{k}_\perp}{ (2 \pi)^{3/2}}~\theta(k^+ )
 ~\left(a(\tilde k)e^{-i\tilde k\cdot \tilde
x}+a^{\dagger}(\tilde
k)e^{i\tilde k\cdot \tilde x}\right) 
\ee 
and
\be
(2\pi)^{3/2}~\sqrt{2\over q^+}~ a(\tilde q)=\int d\tilde x~e^{i\tilde q \cdot \tilde x}~\varphi(\tilde x,0)
\ee
with $q^+\ge 0$.

Combining  the above results, one gets 
\be
\langle \tilde q_2 \tilde q_1 \sigma_1  \big\vert \tilde p,  M,J,J_z;\pi\rangle=
(2\pi)^3 2\sqrt{q^+_1 q^+_2}~\langle 0| a(\tilde q_2) ~
b(\tilde q_1,\sigma_1)  |\tilde p,  M,J^,J_z;\pi\rangle
=
\nonu=q^+_2~\sqrt{2m_F}
\int d\tilde x_2~e^{i\tilde q_2\cdot \tilde x_2}~
\int d\tilde x_1~e^{i\tilde q_1\cdot \tilde x_1}~
\langle 0| \phi(\tilde x_2, 0)~u^{(+)\dagger}(\tilde q_1,\sigma_1)
\psi^{(+)}(\tilde x_1,0)|\tilde p,  M,J,J_z;\pi\rangle
=
\nonu={q^+_2}~\sqrt{m_F\over 2}
\int d\tilde x_2~e^{i\tilde q_2\cdot \tilde x_2}~
\int d\tilde x_1~e^{i\tilde q_1\cdot \tilde x_1}~
\bar u_\alpha(\tilde q_1,\sigma_1)
\gamma^+_{\alpha \beta}~\langle 0| \phi(\tilde x_2, 0)~
\psi_\beta(\tilde x_1,0)|\tilde p,  M,J,J_z;\pi\rangle
\nonu \ee
Finally, by exploiting the translation invariance of the matrix element 
one has
\be
\langle \tilde q_2 \tilde q_1 \sigma_1  \big\vert \tilde p,  M,J,J_z;\pi\rangle=
2 (2\pi)^3~\delta^3(\tilde q_1+\tilde q_2-\tilde p)
~{q^+_2}~\sqrt{m_F\over 2} \nonu \times~
\int d\tilde x~e^{i(\tilde q_1-\tilde q_2)\cdot \tilde x/2}~
\bar u_\alpha(\tilde q_1,\sigma_1)
\gamma^+_{\alpha \beta}~\langle 0| \phi(-\tilde x/2, 0)~
\psi_\beta(\tilde x/2,0)|\tilde p,  M,J,J_z;\pi\rangle
\label{mt2b}
 \ee
Combining  Eqs. \eqref{mt1} and \eqref{mt2b}, one writes the valence wave
function as follows
\be
{p^+}\psi^{J\pi}_{n=2}(q^+_1/p^+ ;{\bf q}_{1\perp};\sigma_1 ;J_z)=
{q^+_2}~\sqrt{m_F\over 2}\int {dx^+\over 2} \delta(x^+/2)
~\int d\tilde x~e^{i(\tilde q_1-\tilde q_2)\cdot \tilde x/2}
\nonu
\times ~
\bar u_\alpha(\tilde q_1,\sigma_1)
\gamma^+_{\alpha \beta}~\langle 0| \phi(-\tilde x/2,-x^+/2)~
\psi_\beta(\tilde x/2,x^+/2)| M,J,J_z;\pi\rangle
=
{q^+_2}~\sqrt{m_F\over 2}\int {dk^-\over 2\pi}
\nonu
\times~
\int d^4x~e^{i k\cdot  x}
~\bar u_\alpha(\tilde q_1,\sigma_1)
\gamma^+_{\alpha \beta}~\langle 0| \phi(-\tilde x/2, -x^+/2)~
\psi_\beta(\tilde x/2,x^+/2)|  M,J,J_z;\pi\rangle
=
\nonu
={q^+_2}~\sqrt{m_F\over 2}\int {dk^-\over 2\pi} ~
\bar u_\alpha(\tilde q_1,\sigma_1)
\gamma^+_{\alpha \beta} \Phi^\pi_\beta(k,p;J_z)
\label{mt3b}\ee
{ where it has been  exploited
\be
\lim_{(x^0+ x^3)\to 0^\pm}~\Bigl[\gamma^+\Bigr]_{\alpha \beta}~\langle 0|T\{\phi(- x/2)~
\psi_\beta( x/2)\} |  M,J,J_z;\pi\rangle=
\nonu
= ~\Bigl[\gamma^+ \Bigr]_{\alpha \beta}~\langle 0|
 \phi(-\tilde x/2,0^\pm)~
\psi_\beta(\tilde x/2,0^\pm)|  M,J,J_z;\pi\rangle
\ee
and non discontinuity in $x^+=0$ has been {\em assumed}.}
After introducing the BS amplitude given in Eq. \eqref{bsa}, one gets the
following expression of  the valence wave function 
 (recall $q^+_1/p^+=\xi$,  ${\bf k}_\perp =({\bf
q}_{1\perp}-{\bf q}_{2\perp})/2$ and ${\bf p}_{\perp}=0$)
\be
\psi^{J\pi}_{n=2}(\xi;{\bf k}_\perp;\sigma_1 ;J_z)=
{ q^+_2\over p^+}~\sqrt{m_F\over 2}\int {dk^-\over 2\pi} 
~\bar u_\alpha(\tilde q_1,\sigma_1)
\gamma^+_{\alpha \beta} \Phi^\pi_\beta(k,p;J_z)=
\nonu
=-{i\over M}~(1-\xi)~\sqrt{m_F\over 2}~\bar u_(\tilde q_1,\sigma_1)\left[
\gamma^+ ~\tilde\phi_1(\xi,\gamma;\kappa^2)+
\gamma^+ {\psla {\bar k}\over M}~\tilde\phi_2(\xi,\gamma;\kappa^2)\right]
{ U}(\tilde p,J_z) 
\label{fock5}\ee
where $\bar  k\equiv \{0, k^+, {\bf k}_\perp\}$.
To achieve the final expression one can use LF spinors, that can be obtained by applying the LF boosts to the
spinors in the CM frame (see for details Ref. \cite{Brod_rev}, where a different
normalization for the LF spinors has been used, i.e. $\bar u u=2m$). Hence,
 one can
rewrite Eq. \eqref{fock5} emphasizing the  contributions where the spins of 
constituent
 and the spin of the system  are {\em aligned} or {\em anti-aligned}. 
The relevant LF spinors are
\be
u(\tilde q_1,\sigma_1)= {1 \over 2 \sqrt{ m_F q^+_1}}~ 
\left[ q^+_1+\beta~m_F +{\bf
k}_\perp\cdot{\bmm \alpha}_\perp\right]~\left(\begin{array} {c} \chi^{\sigma_1} \\
2\sigma_1 \chi^{\sigma_1} \end{array}\right)
\nonu
{ U}(\tilde p,J_z)={1 \over 2M}~ M\left[ 1+\beta\right]~
\left(\begin{array} {c} \chi^{J_z} \\
2J_z \chi^{J_z} \end{array}\right)=  
\left(\begin{array} {c} \chi^{J_z} \\
0 \end{array}\right)
\ee
where $\chi^\sigma$ are the usual two-component spinors. After some lengthy
manipulations, one gets
\be
\psi^{J\pi}_{n=2}(\xi ;{\bf k}_{\perp};\sigma_1 ;J_z)=
-i~(1-\xi)~\sqrt{\xi\over 2M}
\nonu \times ~\left\{ \delta_{\sigma_1,J_z}\left[
\tilde\phi_1(\xi,\gamma;\kappa^2)
-{z\over 2} ~\tilde\phi_2(\xi,\gamma;\kappa^2)\right]
- \delta_{-\sigma_1,J_z}  ~2J_z~
{k_x +  i 2J_z k_y\over M} ~\tilde\phi_2(\xi,\gamma;\kappa^2)
\right\}
\nonu
\label{fock6}
\ee
with $\gamma=|{\bf k}_\perp|^2$. 
\subsection{Valence probability and LF distributions}
\label{vprob}
From   Eq. \eqref{fock6} one can obtain the expression of the valence
probability, given by
\be
P_{val}=  {1 \over (2 \pi)^3}~\sum_{\sigma_1}
\int {d\xi \over 2 ~ \xi (1-\xi)} 
\int
d^2 {\bf k}_{\perp }~
\left| \psi^{J\pi}_{n=2}(\xi  ;{\bf k}_\perp;\sigma_1 ;J_z)\right|^2
=\nonu
={1 \over 4 M(2 \pi)^3}~\int {d\xi ~(1-\xi)} 
\int d^2 {\bf k}_{\perp }~\left[ \left(\tilde\phi_1(\xi,\gamma;\kappa^2)-
{z\over 2}\tilde\phi_2(\xi,\gamma;\kappa^2)\right)^2 +
{|{\bf k}_\perp|^2\over M^2}
\tilde\phi^2_2(\xi,\gamma;\kappa^2)
\right]=
\nonu=
{1 \over 32 M\pi^2}~\int {d\xi ~(1-\xi)} 
\int d{\gamma} ~\left[ \left(\tilde\phi_1(\xi,\gamma;\kappa^2)-
{z\over 2}\tilde\phi_2(\xi,\gamma;\kappa^2)\right)^2 +
{\gamma\over M^2}
\tilde\phi^2_2(\xi,\gamma;\kappa^2)
\right]
\ee
The two contributions, from the aligned configuration and the anti-aligned one,
can be easily singled out.

The LF valence distributions describe:  i) the probability distribution to find a
constituent with a given longitudinal fraction $\xi$, and ii) the probability  to 
find a constituent with transverse momentum $\sqrt{\gamma}=|{\bf k}_\perp|$.
They are defined for the  fermionic constituent as follows
\be
\phi^F(\xi)={1 \over 32 M\pi^2}~ ~(1-\xi) 
\int d{\gamma} ~\left[ \left(\tilde\phi_1(\xi,\gamma;\kappa^2)-
{z\over 2}\tilde\phi_2(\xi,\gamma;\kappa^2)\right)^2 +
{\gamma\over M^2}
\tilde\phi^2_2(\xi,\gamma;\kappa^2)
\right]
\nonu
{\cal P}^F(\gamma)={1 \over 32 M\pi^2}~\int {d\xi ~(1-\xi)} 
 ~\left[ \left(\tilde\phi_1(\xi,\gamma;\kappa^2)-
{z\over 2}\tilde\phi_2(\xi,\gamma;\kappa^2)\right)^2 +
{\gamma\over M^2}
\tilde\phi^2_2(\xi,\gamma;\kappa^2)
\right]
\ee
and  are normalized to $P_{val}$.

\bibliography{naka}

\begin{thebibliography}{40}%
\makeatletter
\providecommand \@ifxundefined [1]{%
 \@ifx{#1\undefined}
}%
\providecommand \@ifnum [1]{%
 \ifnum #1\expandafter \@firstoftwo
 \else \expandafter \@secondoftwo
 \fi
}%
\providecommand \@ifx [1]{%
 \ifx #1\expandafter \@firstoftwo
 \else \expandafter \@secondoftwo
 \fi
}%
\providecommand \natexlab [1]{#1}%
\providecommand \enquote  [1]{``#1''}%
\providecommand \bibnamefont  [1]{#1}%
\providecommand \bibfnamefont [1]{#1}%
\providecommand \citenamefont [1]{#1}%
\providecommand \href@noop [0]{\@secondoftwo}%
\providecommand \href [0]{\begingroup \@sanitize@url \@href}%
\providecommand \@href[1]{\@@startlink{#1}\@@href}%
\providecommand \@@href[1]{\endgroup#1\@@endlink}%
\providecommand \@sanitize@url [0]{\catcode `\\12\catcode `\$12\catcode
  `\&12\catcode `\#12\catcode `\^12\catcode `\_12\catcode `\%12\relax}%
\providecommand \@@startlink[1]{}%
\providecommand \@@endlink[0]{}%
\providecommand \url  [0]{\begingroup\@sanitize@url \@url }%
\providecommand \@url [1]{\endgroup\@href {#1}{\urlprefix }}%
\providecommand \urlprefix  [0]{URL }%
\providecommand \Eprint [0]{\href }%
\providecommand \doibase [0]{http://dx.doi.org/}%
\providecommand \selectlanguage [0]{\@gobble}%
\providecommand \bibinfo  [0]{\@secondoftwo}%
\providecommand \bibfield  [0]{\@secondoftwo}%
\providecommand \translation [1]{[#1]}%
\providecommand \BibitemOpen [0]{}%
\providecommand \bibitemStop [0]{}%
\providecommand \bibitemNoStop [0]{.\EOS\space}%
\providecommand \EOS [0]{\spacefactor3000\relax}%
\providecommand \BibitemShut  [1]{\csname bibitem#1\endcsname}%
\let\auto@bib@innerbib\@empty
\bibitem [{\citenamefont {Salpeter}\ and\ \citenamefont {Bethe}(1951)}]{BS51}%
  \BibitemOpen
  \bibfield  {author} {\bibinfo {author} {\bibfnamefont {E.~E.}\ \bibnamefont
  {Salpeter}}\ and\ \bibinfo {author} {\bibfnamefont {H.~A.}\ \bibnamefont
  {Bethe}},\ }\bibfield  {title} {\enquote {\bibinfo {title} {{A Relativistic
  Equation for Bound-State Problems}},}\ }\href {\doibase
  10.1103/PhysRev.84.1232} {\bibfield  {journal} {\bibinfo  {journal} {Phys.
  Rev.}\ }\textbf {\bibinfo {volume} {84}},\ \bibinfo {pages} {1232--1242}
  (\bibinfo {year} {1951})}\BibitemShut {NoStop}%
\bibitem [{\citenamefont {Binosi}\ \emph {et~al.}(2016)\citenamefont {Binosi},
  \citenamefont {Chang}, \citenamefont {Papavassiliou}, \citenamefont {Qin},\
  and\ \citenamefont {Roberts}}]{Binosi:2016rxz}%
  \BibitemOpen
  \bibfield  {author} {\bibinfo {author} {\bibfnamefont {D.}~\bibnamefont
  {Binosi}}, \bibinfo {author} {\bibfnamefont {L.}~\bibnamefont {Chang}},
  \bibinfo {author} {\bibfnamefont {J.}~\bibnamefont {Papavassiliou}}, \bibinfo
  {author} {\bibfnamefont {S-X.}\ \bibnamefont {Qin}}, \ and\ \bibinfo {author}
  {\bibfnamefont {C.~D.}\ \bibnamefont {Roberts}},\ }\bibfield  {title}
  {\enquote {\bibinfo {title} {{Symmetry preserving truncations of the gap and
  Bethe-Salpeter equations}},}\ }\href {\doibase 10.1103/PhysRevD.93.096010}
  {\bibfield  {journal} {\bibinfo  {journal} {Phys. Rev. D}\ }\textbf {\bibinfo
  {volume} {93}},\ \bibinfo {pages} {096010} (\bibinfo {year} {2016})},\
  \Eprint {http://arxiv.org/abs/1601.05441} {arXiv:1601.05441 [nucl-th]}
  \BibitemShut {NoStop}%
\bibitem [{\citenamefont {Roberts}\ and\ \citenamefont
  {Williams}(1994)}]{Roberts:1994dr}%
  \BibitemOpen
  \bibfield  {author} {\bibinfo {author} {\bibfnamefont {C.~D.}\ \bibnamefont
  {Roberts}}\ and\ \bibinfo {author} {\bibfnamefont {A.~G.}\ \bibnamefont
  {Williams}},\ }\bibfield  {title} {\enquote {\bibinfo {title}
  {{Dyson-Schwinger equations and their application to hadronic physics}},}\
  }\href {\doibase 10.1016/0146-6410(94)90049-3} {\bibfield  {journal}
  {\bibinfo  {journal} {Prog. Part. Nucl. Phys.}\ }\textbf {\bibinfo {volume}
  {33}},\ \bibinfo {pages} {477--575} (\bibinfo {year} {1994})},\ \Eprint
  {http://arxiv.org/abs/hep-ph/9403224} {arXiv:hep-ph/9403224 [hep-ph]}
  \BibitemShut {NoStop}%
\bibitem [{\citenamefont {Alkofer}\ and\ \citenamefont {von
  Smekal}(2001)}]{Alkofer:2000wg}%
  \BibitemOpen
  \bibfield  {author} {\bibinfo {author} {\bibfnamefont {R.}~\bibnamefont
  {Alkofer}}\ and\ \bibinfo {author} {\bibfnamefont {L.}~\bibnamefont {von
  Smekal}},\ }\bibfield  {title} {\enquote {\bibinfo {title} {{The Infrared
  behavior of QCD Green's functions: Confinement dynamical symmetry breaking,
  and hadrons as relativistic bound states}},}\ }\href {\doibase
  10.1016/S0370-1573(01)00010-2} {\bibfield  {journal} {\bibinfo  {journal}
  {Phys. Rept.}\ }\textbf {\bibinfo {volume} {353}},\ \bibinfo {pages} {281}
  (\bibinfo {year} {2001})},\ \Eprint {http://arxiv.org/abs/hep-ph/0007355}
  {arXiv:hep-ph/0007355 [hep-ph]} \BibitemShut {NoStop}%
\bibitem [{\citenamefont {Maris}\ and\ \citenamefont
  {Roberts}(2003)}]{Maris:2003vk}%
  \BibitemOpen
  \bibfield  {author} {\bibinfo {author} {\bibfnamefont {P.}~\bibnamefont
  {Maris}}\ and\ \bibinfo {author} {\bibfnamefont {C.~D.}\ \bibnamefont
  {Roberts}},\ }\bibfield  {title} {\enquote {\bibinfo {title}
  {{Dyson-Schwinger equations: A Tool for hadron physics}},}\ }\href {\doibase
  10.1142/S0218301303001326} {\bibfield  {journal} {\bibinfo  {journal} {Int.
  J. Mod. Phys.}\ }\textbf {\bibinfo {volume} {E12}},\ \bibinfo {pages}
  {297--365} (\bibinfo {year} {2003})},\ \Eprint
  {http://arxiv.org/abs/nucl-th/0301049} {arXiv:nucl-th/0301049 [nucl-th]}
  \BibitemShut {NoStop}%
\bibitem [{\citenamefont {Tandy}(2003)}]{Tandy:2003hn}%
  \BibitemOpen
  \bibfield  {author} {\bibinfo {author} {\bibfnamefont {P.~C.}\ \bibnamefont
  {Tandy}},\ }\bibfield  {title} {\enquote {\bibinfo {title} {{Covariant QCD
  modeling of light meson physics}},}\ }\bibfield  {booktitle} {\emph {\bibinfo
  {booktitle} {{Quarks in hadrons and nuclei. Proceedings, International School
  of Nuclear Physics, 24th Course, Erice, Italy, September 16-24, 2002}}},\
  }\href {\doibase 10.1016/S0146-6410(03)00024-3} {\bibfield  {journal}
  {\bibinfo  {journal} {Prog. Part. Nucl. Phys.}\ }\textbf {\bibinfo {volume}
  {50}},\ \bibinfo {pages} {305--315} (\bibinfo {year} {2003})},\ \bibinfo
  {note} {[,305(2003)]},\ \Eprint {http://arxiv.org/abs/nucl-th/0301040}
  {arXiv:nucl-th/0301040 [nucl-th]} \BibitemShut {NoStop}%
\bibitem [{\citenamefont {Fischer}(2006)}]{Fischer:2006ub}%
  \BibitemOpen
  \bibfield  {author} {\bibinfo {author} {\bibfnamefont {C.~S.}\ \bibnamefont
  {Fischer}},\ }\bibfield  {title} {\enquote {\bibinfo {title} {{Infrared
  properties of QCD from Dyson-Schwinger equations}},}\ }\href {\doibase
  10.1088/0954-3899/32/8/R02} {\bibfield  {journal} {\bibinfo  {journal} {J.
  Phys.}\ }\textbf {\bibinfo {volume} {G32}},\ \bibinfo {pages} {R253--R291}
  (\bibinfo {year} {2006})},\ \Eprint {http://arxiv.org/abs/hep-ph/0605173}
  {arXiv:hep-ph/0605173 [hep-ph]} \BibitemShut {NoStop}%
\bibitem [{\citenamefont {Bashir}\ \emph {et~al.}(2012)\citenamefont {Bashir},
  \citenamefont {Chang}, \citenamefont {Cloet}, \citenamefont {El-Bennich},
  \citenamefont {Liu}, \citenamefont {Roberts},\ and\ \citenamefont
  {Tandy}}]{Bashir:2012fs}%
  \BibitemOpen
  \bibfield  {author} {\bibinfo {author} {\bibfnamefont {A.}~\bibnamefont
  {Bashir}}, \bibinfo {author} {\bibfnamefont {L.}~\bibnamefont {Chang}},
  \bibinfo {author} {\bibfnamefont {I.~C.}\ \bibnamefont {Cloet}}, \bibinfo
  {author} {\bibfnamefont {B.}~\bibnamefont {El-Bennich}}, \bibinfo {author}
  {\bibfnamefont {Y-X}\ \bibnamefont {Liu}}, \bibinfo {author} {\bibfnamefont
  {C.~D.}\ \bibnamefont {Roberts}}, \ and\ \bibinfo {author} {\bibfnamefont
  {P.~C.}\ \bibnamefont {Tandy}},\ }\bibfield  {title} {\enquote {\bibinfo
  {title} {{Collective perspective on advances in Dyson-Schwinger Equation
  QCD}},}\ }\href {\doibase 10.1088/0253-6102/58/1/16} {\bibfield  {journal}
  {\bibinfo  {journal} {Commun. Theor. Phys.}\ }\textbf {\bibinfo {volume}
  {58}},\ \bibinfo {pages} {79--134} (\bibinfo {year} {2012})},\ \Eprint
  {http://arxiv.org/abs/1201.3366} {arXiv:1201.3366 [nucl-th]} \BibitemShut
  {NoStop}%
\bibitem [{\citenamefont {Eichmann}\ \emph {et~al.}(2016)\citenamefont
  {Eichmann}, \citenamefont {Sanchis-Alepuz}, \citenamefont {Williams},
  \citenamefont {Alkofer},\ and\ \citenamefont {Fischer}}]{Eichmann:2016yit}%
  \BibitemOpen
  \bibfield  {author} {\bibinfo {author} {\bibfnamefont {G.}~\bibnamefont
  {Eichmann}}, \bibinfo {author} {\bibfnamefont {H.}~\bibnamefont
  {Sanchis-Alepuz}}, \bibinfo {author} {\bibfnamefont {R.}~\bibnamefont
  {Williams}}, \bibinfo {author} {\bibfnamefont {R.}~\bibnamefont {Alkofer}}, \
  and\ \bibinfo {author} {\bibfnamefont {C.~S.}\ \bibnamefont {Fischer}},\
  }\bibfield  {title} {\enquote {\bibinfo {title} {{Baryons as relativistic
  three-quark bound states}},}\ }\href {\doibase 10.1016/j.ppnp.2016.07.001}
  {\bibfield  {journal} {\bibinfo  {journal} {Prog. Part. Nucl. Phys.}\
  }\textbf {\bibinfo {volume} {91}},\ \bibinfo {pages} {1--100} (\bibinfo
  {year} {2016})},\ \Eprint {http://arxiv.org/abs/1606.09602} {arXiv:1606.09602
  [hep-ph]} \BibitemShut {NoStop}%
\bibitem [{\citenamefont {Sanchis-Alepuz}\ and\ \citenamefont
  {Williams}(2018)}]{Sanchis-Alepuz:2017jjd}%
  \BibitemOpen
  \bibfield  {author} {\bibinfo {author} {\bibfnamefont {H.}~\bibnamefont
  {Sanchis-Alepuz}}\ and\ \bibinfo {author} {\bibfnamefont {R.}~\bibnamefont
  {Williams}},\ }\bibfield  {title} {\enquote {\bibinfo {title} {{Recent
  developments in bound-state calculations using the Dyson-Schwinger and
  Bethe-Salpeter equations}},}\ }\href {\doibase 10.1016/j.cpc.2018.05.020}
  {\bibfield  {journal} {\bibinfo  {journal} {Comput. Phys. Commun.}\ }\textbf
  {\bibinfo {volume} {232}},\ \bibinfo {pages} {1--21} (\bibinfo {year}
  {2018})},\ \Eprint {http://arxiv.org/abs/1710.04903} {arXiv:1710.04903
  [hep-ph]} \BibitemShut {NoStop}%
\bibitem [{\citenamefont {Sauli}(2003)}]{Sauli:2002tk}%
  \BibitemOpen
  \bibfield  {author} {\bibinfo {author} {\bibfnamefont {V.}~\bibnamefont
  {Sauli}},\ }\bibfield  {title} {\enquote {\bibinfo {title} {{Minkowski
  solution of Dyson-Schwinger equations in momentum subtraction scheme}},}\
  }\href {\doibase 10.1088/1126-6708/2003/02/001} {\bibfield  {journal}
  {\bibinfo  {journal} {JHEP}\ }\textbf {\bibinfo {volume} {02}},\ \bibinfo
  {pages} {001} (\bibinfo {year} {2003})},\ \Eprint
  {http://arxiv.org/abs/hep-ph/0209046} {arXiv:hep-ph/0209046 [hep-ph]}
  \BibitemShut {NoStop}%
\bibitem [{\citenamefont {Mezrag}(2019)}]{Mezrag}%
  \BibitemOpen
  \bibfield  {author} {\bibinfo {author} {\bibfnamefont {C.}~\bibnamefont
  {Mezrag}},\ }\bibfield  {title} {\enquote {\bibinfo {title} {private
  communication},}\ }\href@noop {} {\  (\bibinfo {year} {2019})}\BibitemShut
  {NoStop}%
\bibitem [{\citenamefont {Mello}\ \emph {et~al.}(2017)\citenamefont {Mello},
  \citenamefont {de~Melo},\ and\ \citenamefont {Frederico}}]{Mello:2017mor}%
  \BibitemOpen
  \bibfield  {author} {\bibinfo {author} {\bibfnamefont {C.~S.}\ \bibnamefont
  {Mello}}, \bibinfo {author} {\bibfnamefont {J.~P. B.~C.}\ \bibnamefont
  {de~Melo}}, \ and\ \bibinfo {author} {\bibfnamefont {T.}~\bibnamefont
  {Frederico}},\ }\bibfield  {title} {\enquote {\bibinfo {title} {{Minkowski
  space pion model inspired by lattice QCD running quark mass}},}\ }\href
  {\doibase 10.1016/j.physletb.2016.12.058} {\bibfield  {journal} {\bibinfo
  {journal} {Phys. Lett. B}\ }\textbf {\bibinfo {volume} {766}},\ \bibinfo
  {pages} {86--93} (\bibinfo {year} {2017})}\BibitemShut {NoStop}%
\bibitem [{\citenamefont {Nakanishi}(1971)}]{nak71}%
  \BibitemOpen
  \bibfield  {author} {\bibinfo {author} {\bibfnamefont {N.}~\bibnamefont
  {Nakanishi}},\ }\href@noop {} {\emph {\bibinfo {title} {{Graph Theory and
  Feynman Integrals}}}}\ (\bibinfo  {publisher} {Gordon and Breach},\ \bibinfo
  {address} {New York},\ \bibinfo {year} {1971})\BibitemShut {NoStop}%
\bibitem [{\citenamefont {Frederico}\ \emph {et~al.}(2012)\citenamefont
  {Frederico}, \citenamefont {Salm\`e},\ and\ \citenamefont {Viviani}}]{FSV1}%
  \BibitemOpen
  \bibfield  {author} {\bibinfo {author} {\bibfnamefont {T.}~\bibnamefont
  {Frederico}}, \bibinfo {author} {\bibfnamefont {G.}~\bibnamefont {Salm\`e}},
  \ and\ \bibinfo {author} {\bibfnamefont {M.}~\bibnamefont {Viviani}},\
  }\bibfield  {title} {\enquote {\bibinfo {title} {{Two-body scattering states
  in Minkowski space and the Nakanishi integral representation onto the null
  plane}},}\ }\href {\doibase 10.1103/PhysRevD.85.036009} {\bibfield  {journal}
  {\bibinfo  {journal} {Phys. Rev. D}\ }\textbf {\bibinfo {volume} {85}},\
  \bibinfo {pages} {036009} (\bibinfo {year} {2012})}\BibitemShut {NoStop}%
\bibitem [{\citenamefont {Karmanov}\ and\ \citenamefont
  {Carbonell}(2006{\natexlab{a}})}]{CK2006}%
  \BibitemOpen
  \bibfield  {author} {\bibinfo {author} {\bibfnamefont {V.~A.}\ \bibnamefont
  {Karmanov}}\ and\ \bibinfo {author} {\bibfnamefont {J.}~\bibnamefont
  {Carbonell}},\ }\bibfield  {title} {\enquote {\bibinfo {title} {{Solving
  Bethe-Salpeter equation in Minkowski space}},}\ }\href {\doibase
  10.1140/epja/i2005-10193-0} {\bibfield  {journal} {\bibinfo  {journal} {Eur.
  Phys. J. A}\ }\textbf {\bibinfo {volume} {27}},\ \bibinfo {pages} {1--9}
  (\bibinfo {year} {2006}{\natexlab{a}})},\ \Eprint
  {http://arxiv.org/abs/hep-th/0505261} {arXiv:hep-th/0505261 [hep-th]}
  \BibitemShut {NoStop}%
\bibitem [{\citenamefont {Carbonell}\ and\ \citenamefont
  {Karmanov}(2006)}]{CK2006b}%
  \BibitemOpen
  \bibfield  {author} {\bibinfo {author} {\bibfnamefont {J.}~\bibnamefont
  {Carbonell}}\ and\ \bibinfo {author} {\bibfnamefont {V.~A.}\ \bibnamefont
  {Karmanov}},\ }\bibfield  {title} {\enquote {\bibinfo {title} {{Cross-ladder
  effects in Bethe-Salpeter and light-front equations}},}\ }\href {\doibase
  10.1140/epja/i2005-10194-y} {\bibfield  {journal} {\bibinfo  {journal} {Eur.
  Phys. J. A}\ }\textbf {\bibinfo {volume} {27}},\ \bibinfo {pages} {11--21}
  (\bibinfo {year} {2006})},\ \Eprint {http://arxiv.org/abs/hep-th/0505262}
  {arXiv:hep-th/0505262 [hep-th]} \BibitemShut {NoStop}%
\bibitem [{\citenamefont {Carbonell}\ and\ \citenamefont
  {Karmanov}(2010)}]{CK2010}%
  \BibitemOpen
  \bibfield  {author} {\bibinfo {author} {\bibfnamefont {J.}~\bibnamefont
  {Carbonell}}\ and\ \bibinfo {author} {\bibfnamefont {V.~A.}\ \bibnamefont
  {Karmanov}},\ }\bibfield  {title} {\enquote {\bibinfo {title} {{Solving
  Bethe-Salpeter equation for two fermions in Minkowski space}},}\ }\href
  {\doibase 10.1140/epja/i2010-11055-4} {\bibfield  {journal} {\bibinfo
  {journal} {Eur. Phys. J. A}\ }\textbf {\bibinfo {volume} {46}},\ \bibinfo
  {pages} {387--397} (\bibinfo {year} {2010})},\ \Eprint
  {http://arxiv.org/abs/1010.4640} {arXiv:1010.4640 [hep-ph]} \BibitemShut
  {NoStop}%
\bibitem [{\citenamefont {Frederico}\ \emph {et~al.}(2014)\citenamefont
  {Frederico}, \citenamefont {Salm\`e},\ and\ \citenamefont {Viviani}}]{FSV2}%
  \BibitemOpen
  \bibfield  {author} {\bibinfo {author} {\bibfnamefont {T.}~\bibnamefont
  {Frederico}}, \bibinfo {author} {\bibfnamefont {G.}~\bibnamefont {Salm\`e}},
  \ and\ \bibinfo {author} {\bibfnamefont {M.}~\bibnamefont {Viviani}},\
  }\bibfield  {title} {\enquote {\bibinfo {title} {{Quantitative studies of the
  homogeneous Bethe-Salpeter equation in Minkowski space}},}\ }\href {\doibase
  10.1103/PhysRevD.89.016010} {\bibfield  {journal} {\bibinfo  {journal} {Phys.
  Rev. D}\ }\textbf {\bibinfo {volume} {89}},\ \bibinfo {pages} {016010}
  (\bibinfo {year} {2014})}\BibitemShut {NoStop}%
\bibitem [{\citenamefont {Frederico}\ \emph {et~al.}(2015)\citenamefont
  {Frederico}, \citenamefont {Salm{\`e}},\ and\ \citenamefont
  {Viviani}}]{FSV3}%
  \BibitemOpen
  \bibfield  {author} {\bibinfo {author} {\bibfnamefont {T.}~\bibnamefont
  {Frederico}}, \bibinfo {author} {\bibfnamefont {G.}~\bibnamefont
  {Salm{\`e}}}, \ and\ \bibinfo {author} {\bibfnamefont {M.}~\bibnamefont
  {Viviani}},\ }\bibfield  {title} {\enquote {\bibinfo {title} {{Solving the
  inhomogeneous Bethe--Salpeter equation in Minkowski space: the zero-energy
  limit}},}\ }\href {\doibase 10.1140/epjc/s10052-015-3616-1} {\bibfield
  {journal} {\bibinfo  {journal} {Eur. Phys. Jou. C}\ }\textbf {\bibinfo
  {volume} {75}},\ \bibinfo {pages} {398} (\bibinfo {year} {2015})}\BibitemShut
  {NoStop}%
\bibitem [{\citenamefont {Gutierrez}\ \emph {et~al.}(2016)\citenamefont
  {Gutierrez}, \citenamefont {Gigante}, \citenamefont {Frederico},
  \citenamefont {Salm\`e}, \citenamefont {Viviani},\ and\ \citenamefont
  {Tomio}}]{Tomio2016}%
  \BibitemOpen
  \bibfield  {author} {\bibinfo {author} {\bibfnamefont {C.}~\bibnamefont
  {Gutierrez}}, \bibinfo {author} {\bibfnamefont {V.}~\bibnamefont {Gigante}},
  \bibinfo {author} {\bibfnamefont {T.}~\bibnamefont {Frederico}}, \bibinfo
  {author} {\bibfnamefont {G.}~\bibnamefont {Salm\`e}}, \bibinfo {author}
  {\bibfnamefont {M.}~\bibnamefont {Viviani}}, \ and\ \bibinfo {author}
  {\bibfnamefont {Lauro}\ \bibnamefont {Tomio}},\ }\bibfield  {title} {\enquote
  {\bibinfo {title} {{Bethe-Salpeter bound-state structure in Minkowski
  space}},}\ }\href {\doibase 10.1016/j.physletb.2016.05.066} {\bibfield
  {journal} {\bibinfo  {journal} {Phys. Lett. B}\ }\textbf {\bibinfo {volume}
  {759}},\ \bibinfo {pages} {131--137} (\bibinfo {year} {2016})},\ \Eprint
  {http://arxiv.org/abs/1605.08837} {arXiv:1605.08837 [hep-ph]} \BibitemShut
  {NoStop}%
\bibitem [{\citenamefont {de~Paula}\ \emph {et~al.}(2016)\citenamefont
  {de~Paula}, \citenamefont {Frederico}, \citenamefont {Salm\`e},\ and\
  \citenamefont {Viviani}}]{dFSV1}%
  \BibitemOpen
  \bibfield  {author} {\bibinfo {author} {\bibfnamefont {W.}~\bibnamefont
  {de~Paula}}, \bibinfo {author} {\bibfnamefont {T.}~\bibnamefont {Frederico}},
  \bibinfo {author} {\bibfnamefont {G.}~\bibnamefont {Salm\`e}}, \ and\
  \bibinfo {author} {\bibfnamefont {M.}~\bibnamefont {Viviani}},\ }\bibfield
  {title} {\enquote {\bibinfo {title} {{Advances in solving the two-fermion
  homogeneous Bethe-Salpeter equation in Minkowski space}},}\ }\href {\doibase
  10.1103/PhysRevD.94.071901} {\bibfield  {journal} {\bibinfo  {journal} {Phys.
  Rev. D}\ }\textbf {\bibinfo {volume} {94}},\ \bibinfo {pages} {071901}
  (\bibinfo {year} {2016})}\BibitemShut {NoStop}%
\bibitem [{\citenamefont {de~Paula}\ \emph {et~al.}(2017)\citenamefont
  {de~Paula}, \citenamefont {Frederico}, \citenamefont {Salm\`e}, \citenamefont
  {Viviani},\ and\ \citenamefont {Pimentel}}]{dFSV2}%
  \BibitemOpen
  \bibfield  {author} {\bibinfo {author} {\bibfnamefont {W.}~\bibnamefont
  {de~Paula}}, \bibinfo {author} {\bibfnamefont {T.}~\bibnamefont {Frederico}},
  \bibinfo {author} {\bibfnamefont {G.}~\bibnamefont {Salm\`e}}, \bibinfo
  {author} {\bibfnamefont {M.}~\bibnamefont {Viviani}}, \ and\ \bibinfo
  {author} {\bibfnamefont {R.}~\bibnamefont {Pimentel}},\ }\bibfield  {title}
  {\enquote {\bibinfo {title} {{Fermionic bound states in Minkowski-space:
  Light-cone singularities and structure}},}\ }\href {\doibase
  10.1140/epjc/s10052-017-5351-2} {\bibfield  {journal} {\bibinfo  {journal}
  {Eur. Phys. Jou. C}\ }\textbf {\bibinfo {volume} {77}},\ \bibinfo {pages}
  {764} (\bibinfo {year} {2017})},\ \Eprint {http://arxiv.org/abs/1707.06946}
  {arXiv:1707.06946 [hep-ph]} \BibitemShut {NoStop}%
\bibitem [{\citenamefont {Karmanov}\ and\ \citenamefont
  {Carbonell}(2006{\natexlab{b}})}]{karmanov2006bethe}%
  \BibitemOpen
  \bibfield  {author} {\bibinfo {author} {\bibfnamefont {V.A.}\ \bibnamefont
  {Karmanov}}\ and\ \bibinfo {author} {\bibfnamefont {J.}~\bibnamefont
  {Carbonell}},\ }\bibfield  {title} {\enquote {\bibinfo {title}
  {{Bethe-Salpeter equation in Minkowski space with cross-ladder kernel}},}\
  }\href@noop {} {\bibfield  {journal} {\bibinfo  {journal} {Nuclear Physics
  B-Proceedings Supplements}\ }\textbf {\bibinfo {volume} {161}},\ \bibinfo
  {pages} {123--129} (\bibinfo {year} {2006}{\natexlab{b}})}\BibitemShut
  {NoStop}%
\bibitem [{\citenamefont {Gigante}\ \emph {et~al.}(2017)\citenamefont
  {Gigante}, \citenamefont {Nogueira}, \citenamefont {Ydrefors}, \citenamefont
  {Gutierrez}, \citenamefont {Karmanov},\ and\ \citenamefont
  {Frederico}}]{gigante2017bound}%
  \BibitemOpen
  \bibfield  {author} {\bibinfo {author} {\bibfnamefont {V.}~\bibnamefont
  {Gigante}}, \bibinfo {author} {\bibfnamefont {J.~H.~Alvarenga}\ \bibnamefont
  {Nogueira}}, \bibinfo {author} {\bibfnamefont {E.}~\bibnamefont {Ydrefors}},
  \bibinfo {author} {\bibfnamefont {C.}~\bibnamefont {Gutierrez}}, \bibinfo
  {author} {\bibfnamefont {V.~A.}\ \bibnamefont {Karmanov}}, \ and\ \bibinfo
  {author} {\bibfnamefont {T.}~\bibnamefont {Frederico}},\ }\bibfield  {title}
  {\enquote {\bibinfo {title} {{Bound state structure and electromagnetic form
  factor beyond the ladder approximation}},}\ }\href {\doibase
  10.1103/PhysRevD.95.056012} {\bibfield  {journal} {\bibinfo  {journal} {Phys.
  Rev. D}\ }\textbf {\bibinfo {volume} {95}},\ \bibinfo {pages} {056012}
  (\bibinfo {year} {2017})},\ \Eprint {http://arxiv.org/abs/1611.03773}
  {arXiv:1611.03773 [hep-ph]} \BibitemShut {NoStop}%
\bibitem [{\citenamefont {Carbonell}\ \emph {et~al.}(2017)\citenamefont
  {Carbonell}, \citenamefont {Frederico},\ and\ \citenamefont
  {Karmanov}}]{Carbonell:2017kqa}%
  \BibitemOpen
  \bibfield  {author} {\bibinfo {author} {\bibfnamefont {J.}~\bibnamefont
  {Carbonell}}, \bibinfo {author} {\bibfnamefont {T.}~\bibnamefont
  {Frederico}}, \ and\ \bibinfo {author} {\bibfnamefont {V.~A.}\ \bibnamefont
  {Karmanov}},\ }\bibfield  {title} {\enquote {\bibinfo {title} {{Bound state
  equation for the Nakanishi weight function}},}\ }\href {\doibase
  10.1016/j.physletb.2017.04.016} {\bibfield  {journal} {\bibinfo  {journal}
  {Phys. Lett. B}\ }\textbf {\bibinfo {volume} {769}},\ \bibinfo {pages}
  {418--423} (\bibinfo {year} {2017})},\ \Eprint
  {http://arxiv.org/abs/1704.04160} {arXiv:1704.04160 [hep-ph]} \BibitemShut
  {NoStop}%
\bibitem [{\citenamefont {Chang}\ \emph {et~al.}(2013)\citenamefont {Chang},
  \citenamefont {Cloet}, \citenamefont {Cobos-Martinez}, \citenamefont
  {Roberts}, \citenamefont {Schmidt},\ and\ \citenamefont
  {Tandy}}]{Chang:2013pq}%
  \BibitemOpen
  \bibfield  {author} {\bibinfo {author} {\bibfnamefont {L.}~\bibnamefont
  {Chang}}, \bibinfo {author} {\bibfnamefont {I.~C.}\ \bibnamefont {Cloet}},
  \bibinfo {author} {\bibfnamefont {J.~J.}\ \bibnamefont {Cobos-Martinez}},
  \bibinfo {author} {\bibfnamefont {C.~D.}\ \bibnamefont {Roberts}}, \bibinfo
  {author} {\bibfnamefont {S.~M.}\ \bibnamefont {Schmidt}}, \ and\ \bibinfo
  {author} {\bibfnamefont {P.~C.}\ \bibnamefont {Tandy}},\ }\bibfield  {title}
  {\enquote {\bibinfo {title} {{Imaging dynamical chiral symmetry breaking:
  pion wave function on the light front}},}\ }\href {\doibase
  10.1103/PhysRevLett.110.132001} {\bibfield  {journal} {\bibinfo  {journal}
  {Phys. Rev. Lett.}\ }\textbf {\bibinfo {volume} {110}},\ \bibinfo {pages}
  {132001} (\bibinfo {year} {2013})},\ \Eprint {http://arxiv.org/abs/1301.0324}
  {arXiv:1301.0324 [nucl-th]} \BibitemShut {NoStop}%
\bibitem [{\citenamefont {Gao}\ \emph {et~al.}(2017)\citenamefont {Gao},
  \citenamefont {Chang},\ and\ \citenamefont {Liu}}]{Gao:2016jka}%
  \BibitemOpen
  \bibfield  {author} {\bibinfo {author} {\bibfnamefont {Fei}\ \bibnamefont
  {Gao}}, \bibinfo {author} {\bibfnamefont {Lei}\ \bibnamefont {Chang}}, \ and\
  \bibinfo {author} {\bibfnamefont {Yu-xin}\ \bibnamefont {Liu}},\ }\bibfield
  {title} {\enquote {\bibinfo {title} {{Bayesian extraction of the parton
  distribution amplitude from the Bethe-Salpeter wave function}},}\ }\href
  {\doibase 10.1016/j.physletb.2017.04.077} {\bibfield  {journal} {\bibinfo
  {journal} {Phys. Lett.}\ }\textbf {\bibinfo {volume} {B770}},\ \bibinfo
  {pages} {551--555} (\bibinfo {year} {2017})},\ \Eprint
  {http://arxiv.org/abs/1611.03560} {arXiv:1611.03560 [nucl-th]} \BibitemShut
  {NoStop}%
\bibitem [{\citenamefont {Alkofer}\ and\ \citenamefont
  {Alkofer}(2011)}]{Alkofer:2011pe}%
  \BibitemOpen
  \bibfield  {author} {\bibinfo {author} {\bibfnamefont {N.}~\bibnamefont
  {Alkofer}}\ and\ \bibinfo {author} {\bibfnamefont {R.}~\bibnamefont
  {Alkofer}},\ }\bibfield  {title} {\enquote {\bibinfo {title} {{Features of
  ghost-gluon and ghost-quark bound states related to BRST quartets}},}\ }\href
  {\doibase 10.1016/j.physletb.2011.06.073} {\bibfield  {journal} {\bibinfo
  {journal} {Phys. Lett.}\ }\textbf {\bibinfo {volume} {B702}},\ \bibinfo
  {pages} {158--163} (\bibinfo {year} {2011})},\ \Eprint
  {http://arxiv.org/abs/1102.2753} {arXiv:1102.2753 [hep-th]} \BibitemShut
  {NoStop}%
\bibitem [{\citenamefont {Buck}\ \emph {et~al.}(1992)\citenamefont {Buck},
  \citenamefont {Alkofer},\ and\ \citenamefont {Reinhardt}}]{Buck:1992wz}%
  \BibitemOpen
  \bibfield  {author} {\bibinfo {author} {\bibfnamefont {A.}~\bibnamefont
  {Buck}}, \bibinfo {author} {\bibfnamefont {R.}~\bibnamefont {Alkofer}}, \
  and\ \bibinfo {author} {\bibfnamefont {H.}~\bibnamefont {Reinhardt}},\
  }\bibfield  {title} {\enquote {\bibinfo {title} {{Baryons as bound states of
  diquarks and quarks in the Nambu-Jona-Lasinio model}},}\ }\href {\doibase
  10.1016/0370-2693(92)90154-V} {\bibfield  {journal} {\bibinfo  {journal}
  {Phys. Lett.}\ }\textbf {\bibinfo {volume} {B286}},\ \bibinfo {pages}
  {29--35} (\bibinfo {year} {1992})}\BibitemShut {NoStop}%
\bibitem [{\citenamefont {Kusaka}\ \emph {et~al.}(1997)\citenamefont {Kusaka},
  \citenamefont {Simpson},\ and\ \citenamefont {Williams}}]{Kusaka}%
  \BibitemOpen
  \bibfield  {author} {\bibinfo {author} {\bibfnamefont {K.}~\bibnamefont
  {Kusaka}}, \bibinfo {author} {\bibfnamefont {K.}~\bibnamefont {Simpson}}, \
  and\ \bibinfo {author} {\bibfnamefont {A.~G.}\ \bibnamefont {Williams}},\
  }\bibfield  {title} {\enquote {\bibinfo {title} {{Solving the Bethe-Salpeter
  equation for bound states of scalar theories in Minkowski space}},}\ }\href
  {\doibase 10.1103/PhysRevD.56.5071} {\bibfield  {journal} {\bibinfo
  {journal} {Phys. Rev. D}\ }\textbf {\bibinfo {volume} {56}},\ \bibinfo
  {pages} {5071--5085} (\bibinfo {year} {1997})}\BibitemShut {NoStop}%
\bibitem [{\citenamefont {Yan}(1973)}]{Yan}%
  \BibitemOpen
  \bibfield  {author} {\bibinfo {author} {\bibfnamefont {Tung-Mow}\
  \bibnamefont {Yan}},\ }\bibfield  {title} {\enquote {\bibinfo {title}
  {{Quantum Field Theories in the Infinite-Momentum Frame. IV. Scattering
  Matrix of Vector and Dirac Fields and Perturbation Theory}},}\ }\href
  {\doibase 10.1103/PhysRevD.7.1780} {\bibfield  {journal} {\bibinfo  {journal}
  {Phys. Rev. D}\ }\textbf {\bibinfo {volume} {7}},\ \bibinfo {pages}
  {1780--1800} (\bibinfo {year} {1973})}\BibitemShut {NoStop}%
\bibitem [{\citenamefont {Baym}(1960)}]{Baym}%
  \BibitemOpen
  \bibfield  {author} {\bibinfo {author} {\bibfnamefont {Gordon}\ \bibnamefont
  {Baym}},\ }\bibfield  {title} {\enquote {\bibinfo {title} {{Inconsistency of
  Cubic Boson-Boson Interactions}},}\ }\href {\doibase 10.1103/PhysRev.117.886}
  {\bibfield  {journal} {\bibinfo  {journal} {Phys. Rev.}\ }\textbf {\bibinfo
  {volume} {117}},\ \bibinfo {pages} {886--888} (\bibinfo {year}
  {1960})}\BibitemShut {NoStop}%
\bibitem [{\citenamefont {\c{S}avkli}\ \emph {et~al.}(1999)\citenamefont
  {\c{S}avkli}, \citenamefont {Tjon},\ and\ \citenamefont {Gross}}]{Savkli}%
  \BibitemOpen
  \bibfield  {author} {\bibinfo {author} {\bibfnamefont {\c{C}.}\ \bibnamefont
  {\c{S}avkli}}, \bibinfo {author} {\bibfnamefont {J.}~\bibnamefont {Tjon}}, \
  and\ \bibinfo {author} {\bibfnamefont {F.}~\bibnamefont {Gross}},\ }\bibfield
   {title} {\enquote {\bibinfo {title} {{Feynman-Schwinger representation
  approach to nonperturbative physics}},}\ }\href {\doibase
  10.1103/PhysRevC.60.055210, 10.1103/PhysRevC.61.069901} {\bibfield  {journal}
  {\bibinfo  {journal} {Phys. Rev.}\ }\textbf {\bibinfo {volume} {C60}},\
  \bibinfo {pages} {055210} (\bibinfo {year} {1999})},\ \bibinfo {note}
  {[Erratum: Phys. Rev.C61,069901(2000)]},\ \Eprint
  {http://arxiv.org/abs/hep-ph/9906211} {arXiv:hep-ph/9906211 [hep-ph]}
  \BibitemShut {NoStop}%
\bibitem [{\citenamefont {Brodsky}\ \emph {et~al.}(1998)\citenamefont
  {Brodsky}, \citenamefont {Pauli},\ and\ \citenamefont {Pinsky}}]{Brod_rev}%
  \BibitemOpen
  \bibfield  {author} {\bibinfo {author} {\bibfnamefont {S.~J.}\ \bibnamefont
  {Brodsky}}, \bibinfo {author} {\bibfnamefont {H.-C.}\ \bibnamefont {Pauli}},
  \ and\ \bibinfo {author} {\bibfnamefont {S.~S.}\ \bibnamefont {Pinsky}},\
  }\bibfield  {title} {\enquote {\bibinfo {title} {{Quantum chromodynamics and
  other field theories on the light cone}},}\ }\href {\doibase
  10.1016/S0370-1573(97)00089-6} {\bibfield  {journal} {\bibinfo  {journal}
  {Phys. Rept.}\ }\textbf {\bibinfo {volume} {301}},\ \bibinfo {pages}
  {299--486} (\bibinfo {year} {1998})},\ \Eprint
  {http://arxiv.org/abs/hep-ph/9705477} {arXiv:hep-ph/9705477 [hep-ph]}
  \BibitemShut {NoStop}%
\bibitem [{\citenamefont {Link}\ \emph {et~al.}(1991)\citenamefont {Link},
  \citenamefont {Bauer},\ and\ \citenamefont {Ploss}}]{esignal}%
  \BibitemOpen
  \bibfield  {author} {\bibinfo {author} {\bibfnamefont {N.}~\bibnamefont
  {Link}}, \bibinfo {author} {\bibfnamefont {S.}~\bibnamefont {Bauer}}, \ and\
  \bibinfo {author} {\bibfnamefont {B.}~\bibnamefont {Ploss}},\ }\bibfield
  {title} {\enquote {\bibinfo {title} {{Analysis of signals from superposed
  relaxation processes}},}\ }\href {\doibase 10.1063/1.348634} {\bibfield
  {journal} {\bibinfo  {journal} {Jou. Appl. Phys.}\ }\textbf {\bibinfo
  {volume} {69}},\ \bibinfo {pages} {2759--2767} (\bibinfo {year}
  {1991})}\BibitemShut {NoStop}%
\bibitem [{\citenamefont {Dorkin}\ \emph {et~al.}(2008)\citenamefont {Dorkin},
  \citenamefont {Beyer}, \citenamefont {Semikh},\ and\ \citenamefont
  {Kaptari}}]{Dork}%
  \BibitemOpen
  \bibfield  {author} {\bibinfo {author} {\bibfnamefont {S.~M.}\ \bibnamefont
  {Dorkin}}, \bibinfo {author} {\bibfnamefont {M.}~\bibnamefont {Beyer}},
  \bibinfo {author} {\bibfnamefont {S.~S.}\ \bibnamefont {Semikh}}, \ and\
  \bibinfo {author} {\bibfnamefont {L.~P.}\ \bibnamefont {Kaptari}},\
  }\bibfield  {title} {\enquote {\bibinfo {title} {{Two-Fermion Bound States
  within the Bethe-Salpeter Approach}},}\ }\href {\doibase
  10.1007/s00601-008-0196-8} {\bibfield  {journal} {\bibinfo  {journal} {Few
  Body Syst.}\ }\textbf {\bibinfo {volume} {42}},\ \bibinfo {pages} {1--32}
  (\bibinfo {year} {2008})},\ \Eprint {http://arxiv.org/abs/0708.2146}
  {arXiv:0708.2146 [nucl-th]} \BibitemShut {NoStop}%
\bibitem [{\citenamefont {Alvarenga~Nogueira}\ \emph
  {et~al.}(2019)\citenamefont {Alvarenga~Nogueira}, \citenamefont {Frederico},
  \citenamefont {Pace},\ and\ \citenamefont {Salm\`e}}]{scaleinv}%
  \BibitemOpen
  \bibfield  {author} {\bibinfo {author} {\bibfnamefont {J.~H.}\ \bibnamefont
  {Alvarenga~Nogueira}}, \bibinfo {author} {\bibfnamefont {T.}~\bibnamefont
  {Frederico}}, \bibinfo {author} {\bibfnamefont {E.}~\bibnamefont {Pace}}, \
  and\ \bibinfo {author} {\bibfnamefont {G.}~\bibnamefont {Salm\`e}},\
  }\bibfield  {title} {\enquote {\bibinfo {title} {in preparation},}\
  }\href@noop {} {\  (\bibinfo {year} {2019})}\BibitemShut {NoStop}%
\bibitem [{\citenamefont {Chiu}\ and\ \citenamefont {Brodsky}(2017)}]{Chiu}%
  \BibitemOpen
  \bibfield  {author} {\bibinfo {author} {\bibfnamefont {K.~Y.-J.}\
  \bibnamefont {Chiu}}\ and\ \bibinfo {author} {\bibfnamefont {S.~J.}\
  \bibnamefont {Brodsky}},\ }\bibfield  {title} {\enquote {\bibinfo {title}
  {{Angular Momentum Conservation Law in Light-Front Quantum Field Theory}},}\
  }\href {\doibase 10.1103/PhysRevD.95.065035} {\bibfield  {journal} {\bibinfo
  {journal} {Phys. Rev. D}\ }\textbf {\bibinfo {volume} {95}},\ \bibinfo
  {pages} {065035} (\bibinfo {year} {2017})},\ \Eprint
  {http://arxiv.org/abs/1702.01127} {arXiv:1702.01127 [hep-th]} \BibitemShut
  {NoStop}%
\bibitem [{\citenamefont {Luri\'e}\ \emph {et~al.}(1965)\citenamefont
  {Luri\'e}, \citenamefont {Macfarlane},\ and\ \citenamefont
  {Takahashi}}]{Lurie}%
  \BibitemOpen
  \bibfield  {author} {\bibinfo {author} {\bibfnamefont {D.}~\bibnamefont
  {Luri\'e}}, \bibinfo {author} {\bibfnamefont {A.~J.}\ \bibnamefont
  {Macfarlane}}, \ and\ \bibinfo {author} {\bibfnamefont {Y.}~\bibnamefont
  {Takahashi}},\ }\bibfield  {title} {\enquote {\bibinfo {title}
  {{Normalization of Bethe-Salpeter Wave Functions}},}\ }\href {\doibase
  10.1103/PhysRev.140.B1091} {\bibfield  {journal} {\bibinfo  {journal} {Phys.
  Rev.}\ }\textbf {\bibinfo {volume} {40}},\ \bibinfo {pages} {B1091--B1099}
  (\bibinfo {year} {1965})}\BibitemShut {NoStop}%
\end{thebibliography}%

\end{document}